\PassOptionsToPackage{prologue,dvipsnames}{xcolor}
\documentclass[sigconf]{acmart}
\acmConference[ACM CCS 2024]{Proceedings of the 2024 ACM SIGSAC Conference on Computer and Communications Security}{Due 28 Jan 2024}{Salt Lake City, U.S.A.}

\setcopyright{none} 
\acmYear{2024}

\usepackage{comment, graphicx, booktabs, algorithm, algpseudocode, xspace, listings, subcaption, multirow, tikz, siunitx, amsmath, amsthm, xtab}
\usepackage[inline]{enumitem}
\usepackage{array}
\usepackage{threeparttable}
\usepackage{soul}
\usepackage{pbox}
\usepackage[many]{tcolorbox}
\newtcolorbox[auto counter]{mybox}[2][]{enhanced jigsaw,  breakable, #1}

\algnewcommand\algorithmicswitch{\textbf{switch}}
\algnewcommand\algorithmiccase{\textbf{case}}
\algnewcommand\algorithmicassert{\texttt{assert}}
\algnewcommand\Assert[1]{\State \algorithmicassert(#1)}%
\algdef{SE}[SWITCH]{Switch}{EndSwitch}[1]{\algorithmicswitch\ #1\ \algorithmicdo}{\algorithmicend\ \algorithmicswitch}%
\algdef{SE}[CASE]{Case}{EndCase}[1]{\algorithmiccase\ #1}{\algorithmicend\ \algorithmiccase}%
\algtext*{EndSwitch}%
\algtext*{EndCase}%

\graphicspath{{figure/}}

\newcommand{\tool}{\textsc{PromptFuzz}\xspace}
\newcommand*\circled[1]{\tikz[baseline=(char.base)]{
            \node[shape=circle,fill,inner sep=0.8pt] (char) {\textcolor{white}{#1}};}}

\newcommand{\kw}[1]{\textbf{#1}\xspace}
\newcommand{\file}[1]{\textit{#1}\xspace}

\newcommand{\func}[1]{\mbox{\small{\texttt{#1}}}}

\lstdefinelanguage{PromptFuzz}{
  sensitive = true,
  keywords={call, load, update, assert, file},
  morekeywords=[2]={mut*, const*},
  comment=[l]{//},
  morestring=[b]',
  morestring=[b]"
}

\title{Prompt Fuzzing for Fuzz Driver Generation}

\author{Yunlong Lyu}
\affiliation{%
  \institution{Tencent Security Big Data Lab}
  \country{}
}
\email{yunlong.lyu97@gmail.com}

\author{Yuxuan Xie}
\affiliation{%
  \institution{Tencent Security Big Data Lab}
  \country{}
}
\email{yxie0812@gmail.com}

\author{Peng Chen}
\affiliation{%
  \institution{Tencent Security Big Data Lab}
  \country{}
}
\email{spinpx@gmail.com}

\author{Hao Chen}
\affiliation{%
  \institution{University of California, Davis}
  \country{}
}
\email{chen@ucdavis.edu}

\copyrightyear{2024}
\acmYear{2024}
\setcopyright{acmlicensed}
\acmConference[CCS '24]{Proceedings of the 2024 ACM SIGSAC Conference on Computer and Communications Security}{October 14--18, 2024}{Salt Lake City, U.S.A.}
\acmBooktitle{Proceedings of the 2024 ACM SIGSAC Conference on Computer and Communications Security (CCS '24), October 14--18, 2024, Salt Lake City, U.S.A.}
\acmPrice{}
\acmDOI{}
\acmISBN{}

\begin{document}

\begin{abstract}
Crafting high-quality fuzz drivers not only is time-consuming but also requires a deep understanding of the library. However, the state-of-the-art automatic fuzz driver generation techniques fall short of expectations. While fuzz drivers derived from consumer code can reach deep states, they have limited coverage. Conversely, interpretative fuzzing can explore most API calls but requires numerous attempts within a large search space. We propose \tool, a coverage-guided fuzzer for prompt fuzzing that iteratively generates fuzz drivers to explore undiscovered library code. 
To explore API usage in fuzz drivers during prompt fuzzing, we propose several key techniques: instructive program generation, erroneous program validation, coverage-guided prompt mutation, and constrained fuzzer scheduling.
We implemented \tool and evaluated it on 14 real-world libraries. Compared with OSS-Fuzz and Hopper (the
state-of-the-art fuzz driver generation tool), fuzz drivers generated by \tool achieved 1.61 and 1.63 times higher branch coverage than those by OSS-Fuzz and Hopper, respectively. Moreover, the fuzz drivers generated by \tool detected 33 genuine, new bugs out of a total of 49 crashes, out of which 30 bugs have been confirmed by their respective communities.
\end{abstract}



\maketitle

\section{Introduction}
\label{sec:intro}

Fuzzing is crucial for software security and reliability.
\textsf{OSS-Fuzz}~\cite{oss_fuzz}, which deploys state-of-the-art fuzzers for open-source software, has identified and resolved over \num{8900} vulnerabilities and \num{28000} bugs across 850 projects as of February 2023~\cite{oss2023}. These impressive results can largely be attributed to the significant efforts by the contributors to integrate new projects.
When integrating a project for fuzzing, developers select an appropriate fuzzer and write high-quality fuzz drivers. Fuzz drivers are essential because they parse inputs from the fuzzers and invoke the code in the software under test. However, writing high-quality fuzz drivers is challenging because it is both time-consuming and requires a deep understanding of the library. 
Consequently, manually written fuzz drivers often invoke only a small portion of the software's functions and therefore limit the power of fuzz testing~\cite{carpetfuzz, hopper}.

Compared with manually written fuzz drivers, automatic techniques derive fuzz drivers by learning library API usage from either source code or runtime feedback~\cite{fudge, fuzzgen, intelligen, apicraft, winnie, utopia, hopper, afgen, titan_fuzz, oss-llm}. \textsf{FUDGE}~\cite{fudge}, \textsf{FuzzGen}~\cite{fuzzgen}, and \textsf{UTopia}~\cite{utopia} extract the code of API usage from source code statically, while \textsf{APICraft}~\cite{apicraft} and \textsf{WINNIE}~\cite{winnie} record the sequences of API calls from the execution traces of processes dynamically. However, since the traces contain only the API call sequences invoked by the consumer code, this method cannot learn valid API usage that is absent in the consumer code.
\textsf{Hopper}, the state-of-the-art fuzz driver generation solution, transforms the problem of library fuzzing into the problem of interpretative fuzzing, which learns valid API usage from dynamic feedback of API invocations~\cite{hopper}. Although it can cover most API functions, it requires many attempts in a vast search space to find useful API invocation sequences that reach deep states.

Large language models (LLMs) have demonstrated remarkable success in generating program code, promising to effectively explore a wide range of API usage without relying on consumer code. Exemplified by the \textsf{GPT-series}~\cite{gpt2, gpt3, instruct_gpt, gpt4}, they are trained on extensive code corpora and are able to produce code that aligns with user intentions.
Although prior work~\cite{oss-llm,titan_fuzz,zhang2023understanding} attempted to use LLMs for generating fuzz drivers, their instructions for generating fuzz drivers were limited to specific scenarios. As a result, the generated fuzz drivers suffered from low API usage diversity and failed to cover infrequently used code or deep states. 

To address these challenges, we introduce \tool, a coverage-guided fuzzer that iteratively mutates prompts to explore undiscovered library code.
Thanks to coverage guidance, the mutated prompts can direct LLMs to produce code that triggers complex scenarios or enters deep states. Since LLMs sometimes generate erroneous code, \tool employs multiple program error oracles for validating the generated code. The workflow of \tool is as follows:
\begin{enumerate*}
    \item Prompt LLMs with crafted instructions to generate programs focusing on the provided library API functions.
    \item Eliminate erroneous programs which fail to execute or trigger false positives.
    \item Guide the mutation of the LLMs prompts with the feedback of code coverage of the generated programs.
    \item Convert the arguments of library API calls inside the generated programs from constants to variables whose assignments can be mutated during fuzzing.
\end{enumerate*}
Finally, these fuzz drivers are fused into a fuzz driver compatible with existing fuzzers.

We implemented \tool and assessed it on 14 real-world libraries. Compared with \textsf{OSS-Fuzz} and Hopper~\cite{hopper}, the fuzz drivers generated by \tool achieved 1.61 and 1.63 higher branch coverage than that by \textsf{OSS-Fuzz} and \textsf{Hopper}, respectively. Additionally, the fuzz drivers generated by \tool identified 33 genuine, new bugs from 49 crashes, out of which 30 bugs have been confirmed by their respective communities. Moreover, \tool's power scheduling effectively guides LLMs to generate programs that explore deep library code in most libraries.

\section{Background}

\begin{figure}[t]
\begin{lstlisting}[language=C]
#include <vpx/vp8dx.h>
#include <vpx/vp8cx.h>
#include <vpx/vpx_decoder.h>

extern "C" int LLVMFuzzerTestOneInput(const uint8_t *data, size_t size) {
    // Create the decoder configuration
    vpx_codec_dec_cfg_t dec_cfg = {0};
    ...
    // Initialize the decoder
    vpx_codec_ctx_t decoder;
    vpx_codec_iface_t *decoder_iface = vpx_codec_vp8_dx();
    vpx_codec_err_t decoder_init_res = vpx_codec_dec_init_ver(&decoder, decoder_iface, &dec_cfg, 0, VPX_DECODER_ABI_VERSION);
    if (decoder_init_res != VPX_CODEC_OK) {
        return 0;
    }
    // Process the input data
    vpx_codec_err_t decode_res = vpx_codec_decode(&decoder, data, size, NULL, 0);
    if (decode_res != VPX_CODEC_OK) {
        vpx_codec_destroy(&decoder);
        return 0;
    }
    // Get the decoded frame
    vpx_image_t *img = NULL;
    vpx_codec_iter_t iter = NULL;
    while ((img = vpx_codec_get_frame(&decoder, &iter)) != NULL) {
        // Process the frame
        vpx_img_flip(img);
        ...
    }
    // Cleanup
    vpx_codec_destroy(&decoder);
    return 0;
}
\end{lstlisting}
\caption[Fuzz driver example]{A fuzz driver for \textsf{libvpx}}
\label{fig:llm_example}
\end{figure}

\subsection{Library Fuzzing}

Library fuzzing has become increasingly important due to the widespread use of libraries in software development. Unlike command-line programs, which process a byte stream as input, libraries possess multiple access points (i.e., API functions) that require more stringent inputs to adhere to strict format constraints. 
To leverage existing fuzzers~\cite{afl, aflfast, libfuzzer, angora, Chen:2019:Matryoshka}, fuzz drivers are developed to serve as delegates. These drivers accept random bytes from the fuzzer and subsequently convert these bytes into well-structured arguments for API calls.

\autoref{fig:llm_example} depicts a fuzz driver designed for \textit{libvpx}.
This driver fulfills three main functions: \circled{1} it properly invokes API functions to simulate the video decoding process. To ensure proper termination, the code of handling errors and reclaiming resources is incorporated to handle trivial errors; 
\circled{2} it constructs input arguments from randomly generated bytes, which must be carefully formatted due to their complex constraints. For example, the final parameter of \func{vpx\_codec\_dec\_init\_ver} (line 12) accepts only integers within a specified range, as defined by macros, and the third parameter of \func{vpx\_codec\_decode} (line 17) need to correspond to the length of the second parameter; 
\circled{3} it exercises the API functions to reach as much code as possible. Starting from line 25, the loop continuously fetches frames from the decoder and feeds them to the processing API functions in the loop body. In this way, it optimizes the throughput of the byte stream input.

To develop a high-quality fuzz driver, it is essential to adhere to the constraints of the target library and thoroughly test its API functions to achieve comprehensive code coverage. This requires a comprehensive understanding of the target libraries, rendering the automatic generation of fuzz drivers a challenging task.

\subsection{Large Language Model}
Large Language Models (LLMs) are deep learning models with sophisticated architectures and numerous parameters, allowing them to acquire knowledge from vast amounts of textual data.
\textsf{GPT3}~\cite{gpt3}, \textsf{ChatGPT}~\cite{openai_chat} and \textsf{GPT4}~\cite{gpt4}  are current representative examples of LLMs.
LLMs are trained to predict the next word, denoted as $w_{n+1}$, given a sequence of words $w_1, w_2, ..., w_n$, by maximizing the objective function of the language model, as shown in Equation \ref{equ:lm}. 
\begin{equation}
\label{equ:lm}
    P(w_1, w_2, ..., w_n) = \prod_{i=1}^n P(w_i | w_1, w_2, ..., w_{i-1})
\end{equation}
In the inference phase,  LLMs auto-regressively generate the next token, $w_{n+1}$, based on prior tokens $w_1, w_2, ..., w_n$, utilizing the model weights learned from their extensive parameters.
The starting tokens provided by users are known as a \textbf{prompt}.
To ensure that LLMs produce output that is consistent with user instructions and aligns with their intents, a series of LLMs~\cite{openai_chat, gpt4, llama, ernie, bard} have been enhanced with the training of Reinforcement Learning from Human Feedback (\textsf{RLHF})~\cite{instruct_gpt}, such as \textsf{ChatGPT} and \textsf{GPT4}.

\begin{figure*}[t]
  \centering \includegraphics[width=1\linewidth]{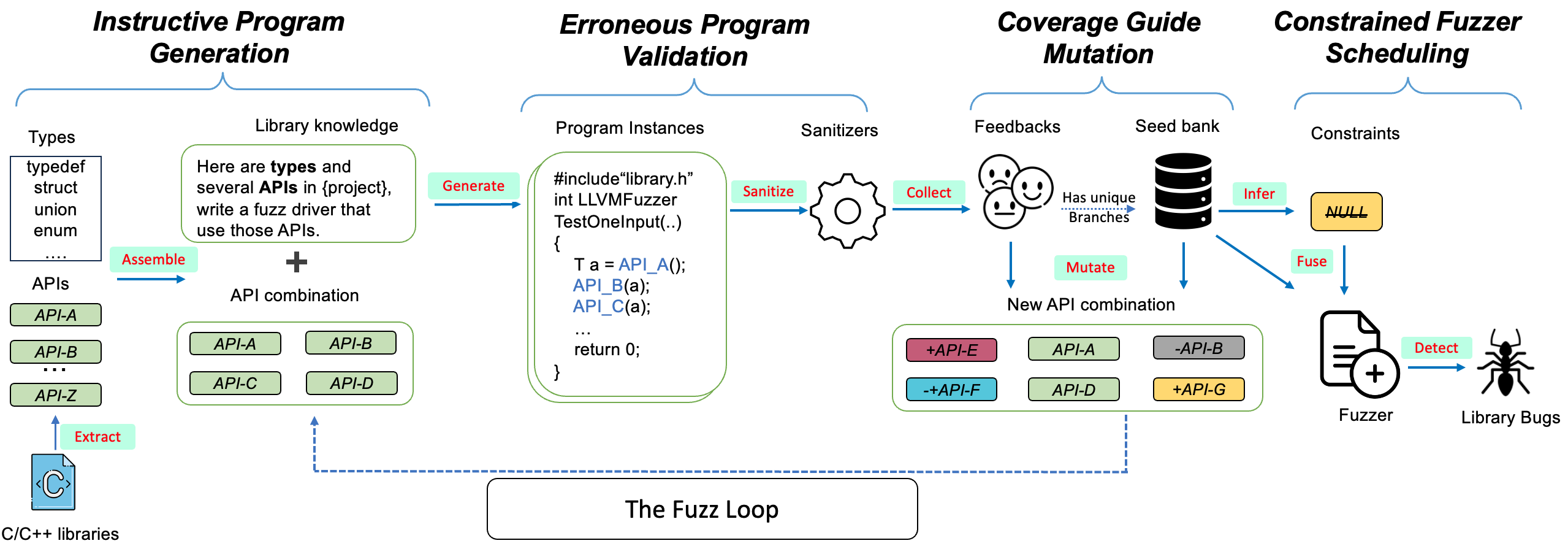}
  \caption[workflow]{Fuzz driver generation in \tool. \textit{Seed} represents a program instance generated by LLMs.}
  \label{fig: workflow}
\end{figure*}

\subsection{LLM-based Fuzz Driver Generation}
Very recently, there has been a growing interest in leveraging LLMs to enhance fuzzing tasks~\cite{xia2023universal, titan_fuzz, fuzzomatic, oss-llm,Zhao:2023}. The primary challenges associated with using LLMs for fuzzing include automatic prompt construction and output validation.

In LLM-based fuzz driver generation, prompts usually consist of task descriptions and contextual information. To be as informative and instructive as possible, the task description should specify at least the target library and the API functions to be included for this round of code generation. Among the early attempts~\cite{titan_fuzz, xia2023universal, llm_edge}, researchers assign only one API for each prompt as the target. However, the code generated with such prompts tends to have the problem of being too simple or demonstrates rather common usage. In contrast, bugs usually exist in corner cases or complicated scenarios.
On the other hand, the code generated by LLMs is hard to directly utilize for fuzzing as it is susceptible to be erroneous~\cite{asleep, lost_at_c, evalplus}. Current works~\cite{fuzzomatic, oss-llm} rely on compilers or simple rules for output validation. However, these ways only report syntactic problems or shallow logic errors and fail to detect complex semantic errors (e.g., incorrect library usage). When utilized as fuzz drivers, such buggy code will incur many false positives.


    

\section{Design}

\subsection{Overview}
\tool generates high-quality fuzz drivers for effective library bug detection via coverage-guided LLM prompt construction.
Unlike grey-box fuzzers, which mutate input bytes to reach deeper program code, \tool mutates LLM prompts to produce programs that cover a broader range of library API utilization.
Initially, \tool constructs a prompt using randomly selected library API functions. Then, it mutates this prompt based on coverage feedback until the fuzzing reaches convergence for the target library. 
The mutations target the API functions within the prompts to generate diverse programs.
Simultaneously, the generated programs are validated at runtime to ensure correctness.
The workflow of \tool is depicted in \autoref{fig: workflow}.
\begin{enumerate}
    \item \tool extracts function signatures and type definitions from the header files of a C/C++ library and uses them to construct prompts for instructing LLMs to generate programs that call these functions.
    \item \tool executes the generated programs, validates them based on their runtime behavior, and eliminates the erroneous ones. \tool also collects code coverage during the executions.
    \item \tool stores programs that pass the validation in a \textit{seed bank}. It then uses their code coverage as feedback to mutate prompts towards API functions that are more likely to explore new code paths. This iterative process continues until \tool discovers no new paths or it exhausts the query budget.
    \item Finally, \tool infers the constraints imposed on library API functions within the seed programs. It converts the arguments of library API calls from constants, which LLMs generated, into variables that can take arbitrary values provided by the fuzzer while preserving the inferred constraints. To detect library bugs, \tool consolidates all the converted seed programs in a fuzz driver and then schedules each seed program to be fuzzed with random bytes from the fuzzers.
\end{enumerate}


\begin{figure}[t]
  \centering \includegraphics[width=1\linewidth]{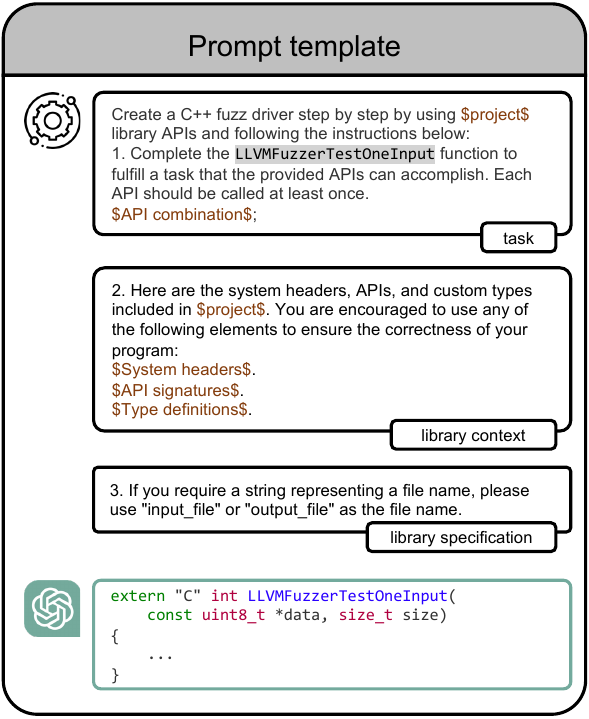}
  \caption[prompt design]{Prompt template}
  \label{fig: prompt}
\end{figure}

\subsection{Instructive Program Generation} \label{sec: generation}
\tool instructs LLMs to generate the desired programs through zero-shot prompting ~\cite{zero_cot}.
It uses LLMs that have been trained on public code datasets and fine-tuned using \textsf{RLHF}~\cite{instruct_gpt} as the generator.
Those LLMs possess the capability to generate code that both conforms to programming syntax and semantics and also aligns with instructions. 
We chose \textsf{ChatGPT}~\cite{openai_chat} and \textsf{GPT-4}~\cite{gpt4} as LLMs. 
While the generated programs may not always strictly follow the instructions, they help explore valid library usage.
Therefore, we use instructions in LLM prompts to steer LLMs toward generating desirable programs for library fuzzing.

\tool constructs prompts that instruct LLMs to generate programs with specific combinations of library API functions. 
To synthesize such an LLM prompt, \tool fills in a prompt template (\autoref{fig: prompt}) with extracted library gadgets and a designated API combination.
For effective fuzz driver generation, a template has the following components:
\begin{itemize}
    \item The \textit{task} component details the intended programs that LLMs should generate. It specifies which API functions from the libraries are mandatory within a \textit{LLVMFuzzerTestOneInput} function. 
    \item The \textit{library context} includes API signatures, custom type definitions, and headers in the library. 
    Given the context length limitations and token cost of current LLMs, \tool restricts the number of API functions and custom types in \textit{library context}.
When a library has too many API functions, exceeding 100 for instance, \tool uses a random selection strategy to choose a manageable subset each time.
For custom types, \tool selects only those types that are used by the chosen API functions for relevance and efficiency.
By integrating a contextual understanding of the libraries, we can significantly reduce the occurrence of "hallucination" code produced by LLMs~\cite{hallucination, bi-etal-2019-incorporating, peng2023check, lewis2020retrieval}.
\item The \textit{library specification} guides LLMs to generate code that adheres to specified patterns required by the libraries. Some library API functions may read inputs from files, file streams, or file descriptors, which may deviate from the standard routines of fuzz drivers. By incorporating the relevant library specifications into the prompts for LLMs, we help LLMs generate code patterns that adhere to these specifications.
\end{itemize}
Once \tool fills in these components to generate a prompt, it queries LLMs with the prompt to generate programs. 

\subsection{Erroneous Program Elimination}\label{sec: elimination}

\begin{figure}[t]
  \centering \includegraphics[width=1\linewidth]{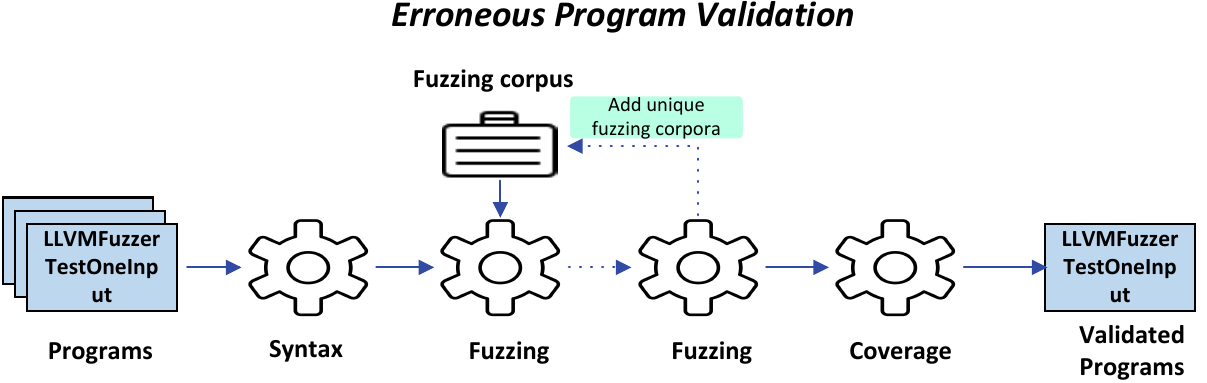}
  \caption[Validation]{Erroneous program validation. 
  \textit{corpora} represents a program input, and \textit{fuzzing corpus} represents a collection of program inputs.}
  \label{fig: elimination}
\end{figure}

Limited by training data bias and imperfect code synthesis ability of LLMs, the code generated by LLMs may be erroneous~\cite{asleep, lost_at_c, evalplus}.
A good target for fuzzing should be, at the very least, free of any errors in the code itself so that all runtime errors are attributed to the library code that the target calls~\cite{good_fuzz_target,zhang2023understanding}. Since LLMs are unable to generate error-free programs consistently, we developed a technique for identifying erroneous programs generated by LLMs.

Identifying syntactic errors is easy, but identifying semantic errors is tricky. In programs for testing libraries, semantic errors include both library misuses and general bugs, such as memory bugs and incorrect control flows. Precisely identifying these errors is challenging because it requires in-depth knowledge of the library and complicated static and dynamic analyses that are likely slow.


Prior attempts to tackle these challenges mainly focused on leveraging existing knowledge to detect buggy code patterns.
Rule-based approaches required manual rules for detection~\cite{avatar, contractBasedApr, jiang2018aprspace}. Learning-based approaches mined correct or incorrect usage from sources, such as consumer code, code patches, or documents~\cite{mutapi, lv2020rtfm, correction, zeng2021mining, Exception, kang2021active, nielebock2022automated, yang2022api}.
Unfortunately, neither approach was effective. First, they, particularly learning-based methods, suffered from poor precision and recall~\cite{amann2018systematic}, resulting in either high false positive rates in detecting library bugs or fuzz drivers with low code coverage. Second, gathering learning materials and writing rules were laborious, and the materials and rules could not be shared across libraries. 



To address these challenges,
\tool eliminates erroneous programs in three steps (\autoref{fig: elimination}):
\begin{enumerate}
    \item It removes programs with syntax errors as identified by C/C++ compilers.
    \item It compiles the remaining programs into executables and incorporates multiple runtime sanitizers, which capture and analyze deviations from expected behavioral patterns.
    \item \tool fuzzes these programs using the provided corpus and removes any program where the sanitizers detect deviations. During fuzzing, \tool adds the inputs that trigger unique behavior to the corpus, expanding the corpus for more thorough runtime-based validation.
    After fuzzing, \tool calculates the code coverage achieved by the programs and removes those that fail to meet the code coverage criteria, indicating the sufficient exercise of the library API functions.
\end{enumerate}
    
\subsubsection{Fuzzing Validation} \label{sec: fuzzing_validation}

To keep as many valid generated programs as possible for fuzzing libraries, \tool conservatively removes only programs exhibiting identifiable errors.
For this purpose, \tool compiles the programs with several runtime error sanitizers --- Address-Sanitizer (\textsf{ASan}~\cite{asan}), Undefined-Behavior-Sanitizer (\textsf{UBSan}~\cite{ubsan}), and File-Sanitizer (see in \autoref{sec:fsan}) --- which can identify violations based on predefined rules with no false positives. 
\tool then executes the programs with a provided corpus, monitors their runtime behavior, and removes the programs where the sanitizers report violations.
To minimize API misuse in the generated programs, \tool does not differentiate whether the reported violations occurred in the generated code or library code because the latter could also be triggered by API misuse. We analyzed this rule and found that it removed many generated programs with API misuses even though it occasionally removed generated programs that triggered true library bugs.

Initially, the program's inputs for fuzzing can be either the seed inputs~\cite{oss_corpus} provided by the developers or simply empty.
These seed inputs typically consist of representative inputs for certain fuzz drivers and therefore serve as a good starting point for fuzzing.
However, as LLMs generate these programs using various API functions to accomplish diverse tasks, as explained in \autoref{sec: generation}, many programs may require custom inputs, so the initial seed inputs may be unsuitable for those programs. This reduces code coverage and leads to undiscovered errors.

To maximize the range of code \tool can examine, \tool continuously evolves the fuzzing corpus throughout the fuzzing process.
After a program passes the previous validation processes, \tool uses a grey-box fuzzer to mutate its inputs while simultaneously monitoring code coverage.
If the code coverage increases within each time interval (e.g., 60 seconds),  \tool continues fuzzing until it exhausts the time budget (e.g., 600 seconds).
Subsequently, \tool adds the inputs that have triggered new code coverage into the fuzzing corpus.
The objective here is to generate specific inputs required by each program, rather than triggering bugs within the library code.
Although short-term fuzzing may not instantaneously produce the required program inputs, \tool is designed to iteratively refine and evolve the fuzzing corpus over subsequent fuzzing rounds, so this continuous evolution improves the likelihood of generating suitable program inputs over time.

To show how the fuzzing process works, we take the program generated by LLMs shown in \autoref{fig:llm_example} as an example. 
This program uses the \textsf{libvpx} API functions to decode a fragment of encoded video frames and processes each frame iteratively. However, the initial seed inputs of \textsf{libvpx} was a collection of video stream files (i.e., IVF) that contained both headers and video frames, which were unsuitable for the scenario depicted in \autoref{fig:llm_example}. When \tool executed the program directly on the initial seed inputs, the call to \func{vpx\_code\_decode()} (line 17) returned a \func{VPX\_CODEC\_UNSUP\_BITSTREAM} error, causing the program to exit immediately at line 20.
The sanitizers failed to report the potential risks in the code at lines 23--32 because of the program's early exit.
After multiple rounds of mutation of the initial fuzzing corpus, \tool generated a suitable input for \func{vpx\_codec\_decode()} that passed the error checking at line 18, which allowed the sanitizers to validate the hitherto unchecked code after this line.

\subsubsection{Coverage Validation}
Fuzzing validation (\autoref{sec: fuzzing_validation}) accurately identifies errors violating sanitizer rules.
However, due to the nature of runtime analysis, it struggles to validate the code that is hard to reach.
For example, in line 17 of \autoref{fig:llm_example}, the code will not execute if the call to \func{vpx\_codec\_dec\_init()} in line 12 contains incorrect arguments, but the sanitizers are unable to detect this API misuse because it does not violate any sanitization rules. To overcome this limitation, \tool validates programs based on their runtime code coverage.

To conduct coverage validation, \tool gathers code coverage for the remaining programs and identifies the \textit{critical paths}, defined as the paths that contain the greatest number of library API calls in the control flow graph of a program. Critical paths represent the API usage that we want to test rather than error-handling code.
For example, the critical path in \autoref{fig:llm_example} executes lines 11, 12, 17, 25, 27, and 31, capturing the essential API calls within the program.
Based on this analysis, \tool removes the programs where an API call on a critical path has not been executed.
This process encourages important API usage to be thoroughly tested.
We concentrate on critical paths rather than all program paths, as certain error-handling code is difficult to access or requires specific configurations.
    
Coverage validation has two benefits. First, prior methods that relied solely on runtime validation could not determine the correctness of unreachable code in programs. Although fuzzing is employed to evolve the fuzzing corpus, it does not guarantee that all programs have been generated with suitable inputs.
Coverage validation removes many programs with unreachable API calls. While this approach may mistakenly exclude programs without API misuse, it significantly reduces false positives in bug detection.
Second, certain library API uses do not trigger abnormal behavior that runtime sanitizers can capture, e.g., the erroneous API initialization in \autoref{fig:llm_example}. Coverage validation can exclude the programs containing these API misuses.
    
\subsection{Coverage-Guide Prompt Mutation}
To create prompts for successive rounds, \tool mutates the API combinations within the previous prompts.
Although LLMs are able to generate programs by combining different API functions, randomly combining them in the prompts would be inefficient.
\tool employs API-level power scheduling and prompt-level mutation strategies to generate effective prompts using code coverage as feedback. 

\subsubsection{Power Schedule}
\tool schedules API functions in LLM prompts based on their energy.
More prompts containing a particular API function often correlate with more code coverage but also raise query costs.
Beyond a certain point,
    if the code coverage for an API function plateaus, the benefit of prompting it diminishes.
Therefore, an effective power scheduling strategy should maximize library code coverage while minimizing the number of LLM queries.
Drawing inspiration from \textsf{AFLFast}\cite{aflfast}, \tool implements monotonous power scheduling, which reduces the energy of well-tested API functions.

Initially, \tool assigns equal energy for each API function and maintains a set of visited branches and call graphs for the library under test.
During each iteration of fuzzing, it updates the visited branches and calculates each API function's branch coverage, as shown in \autoref{equ:func_cov}. When computing an API function's branch coverage, it considers not only the branches within the function's body but also the branches within the bodies of any recursive callees.
\begin{equation}
\label{equ:func_cov}
\mathrm{cov}(i) = \frac{\text{covered branches inside }i}{\text{total branches inside }i}
\end{equation}
For an API function $i$, \tool updates its energy $energy(i)$ following the exponential schedule in \textsf{AFLFast}~\cite{aflfast}, shown in \autoref{equ:energy}:  
\begin{equation}
\label{equ:energy}
    \mathrm{energy}(i) = \frac{1 - \mathrm{cov}(i)}{(1 + \mathrm{seed}(i))^E \times (1 + \mathrm{prompt}(i))^E}
\end{equation}
where $\mathrm{seed}(i)$ is the number of seed programs that call $i$, $\mathrm{prompt}(i)$ is the number of prompts that contain $i$, and $E$ is an exponent to regulate the frequency of $i$. 
As a result, the fewer times \tool has exercised an API function, the higher energy \tool assigns to the function, and the higher probability \tool will include the function in future prompts with.
    
\subsubsection{Mutation strategies}
\tool mutates API function combinations in prompts to instruct program generation.
\tool borrows the following mutation strategies, which are commonly used in traditional fuzzers~\cite{afl, honggfuzz, titan_fuzz, graphfuzz, superion}, but applies them to API functions in prompts:
\begin{itemize}
    \item $\mathrm{Insertion}(C, A)$: Insert API function $A$ into combination \textit{C}.
    \item $\mathrm{Replacement}(C, A, B)$: Replace API function $A$ in combination \textit{C} with API function $B$.
    \item $\mathrm{Crossover}(C, S)$: Merge combinations \textit{C} and \textit{S} into a new combination.
\end{itemize}
Guided by the energy of API functions, \tool schedules the mutations to assemble API function combinations to generate previously unexplored functions.
However, combining API functions based solely on their degrees of exploration without considering their dependencies prevents LLMs from exploring complex API relations.
To overcome this limitation, for each seed program, \tool gathers the following statistics, which reflect the effectiveness of API combinations in prompts:   
\begin{itemize}
    \item Density: The maximum number of library API calls sharing explicit data dependency.
    \item Unique branches: The number of unique branches triggered during program execution.
\end{itemize}
\tool quantifies a program's quality by \autoref{equ:quality}, which assigns a higher quality to the programs that have more correlated API calls and that discover more branches.
\begin{equation}
    \label{equ:quality}
    \mathrm{quality}(g) = \mathrm{density}(g) \times (1 + \mathrm{unique\_branches}(g))
\end{equation}

During each iteration of fuzzing, \tool explores the seed bank and updates the qualities of those seed programs.
Using the feedback from library API energies and seed qualities, \tool applies \autoref{alg:mutation} to select a new API combination to be used in the next iteration.
If the current iteration has insufficient seed programs, \tool enters the warm-up stage (lines 3--7 in \autoref{alg:mutation}), which randomly selects high-energy API functions to explore previously undiscovered library usage.
In the mutation stage (lines 9--23 in \autoref{alg:mutation}), \tool uses the sequence of API calls on the seed program's critical path as the pivot for mutation, where it discards the API calls that do not interact with others. Focusing mutation on the pivot allows \tool to explore intricate API usage.
Finally, \tool uses the new API combination to construct a prompt for the next iteration of program generation.

\begin{algorithm}
    \begin{algorithmic}[1]
    \Function{Mutation}{$APIs, Seeds$} 
    \State $Comb = \{\}$
    \If {$WarmUp(Seeds)$}
        \While {$len(Comb) < DefaultLen$}
            \State $A \gets ChooseByEnerge(APIs)$
            \State $Insert(Comb, A)$
        \EndWhile
        \State \kw{return} $Comb$
    \EndIf 
    \State $seed \gets ChooseSeedByQuality(Seeds)$
    \State $Comb \gets CriticalCalls(seed)$
    \State $mutator \gets RandChoose(Insert, Replace, Crossover)$
      \Switch{$mutator$}
    \Case{$Insert$}
        \State $A \gets ChooseByEnerge(APIs)$
        \State $Insert(Comb, A)$
    \EndCase
    \Case{$Replace$}
        \State $A \gets Choose(Comb)$
        \State $B \gets ChooseByEnerge(APIs)$
        \State $Repalce(Comb, A, B)$
    \EndCase
    \Case{$Crossover$}
        \State $seedB \gets ChooseSeedByPrevious(seed, Seeds)$
        \State $CombB \gets CriticalCalls(seedB)$
        \State $Comb = CrossOver(Comb, CombB)$
    \EndCase
    \EndSwitch
    \State \kw{return} $Comb$
    \EndFunction
    \end{algorithmic}
\caption{Selecting a new API combination}
\label{alg:mutation}
\end{algorithm}

\subsection{Constrained Fuzzer Scheduling}
In the final stage, \tool combines seed programs into a fuzz driver and schedules it to be fuzzed.
We mark the seeds that trigger unique branches as \textit{unique seeds}.
Since the \textit{unique seeds} encompass nearly all discovered API functions, only they are included in the fuzz driver. 

To empower seed programs for fuzzing, \tool infers the constraints on API function arguments from the programs stored in the seed bank. Under these constraints, \tool instruments the \textit{unique seeds} by converting their API function arguments from constants to variables capable of receiving arbitrary bytes from the fuzzers. Finally, \tool integrates these seeds into a single fuzz driver, which schedules each seed randomly.

\subsubsection{Argument Constraint Inference}\label{sec:infer}
\tool infers the argument constraints of API functions via statistical inference.
For arguments of an immutable array or scalar type, \tool infers them as potential recipients for random bytes from the fuzzers. These arguments are commonly subject to the following constraints, which can significantly affect the effectiveness of fuzz drivers~\cite{hopper, utopia}:
\begin{itemize}
    \item $\mathrm{ArrayLength}(A, n)$: $n$ is the capacity of the array $A$.
    \item $\mathrm{ArrayIndex}(A, i)$: $i$ is an index to the array $A$.
    \item $\mathrm{FileName}(S)$: $S$ is a file path.
    \item $\mathrm{FormatString}(S)$: $S$ is a format string.
    \item $\mathrm{AllocSize}(n)$: $n$ is the size of buffer allocation.
    \item $\mathrm{FileDesc}(fd)$: $fd$ is a file descriptor.
\end{itemize}

\tool infers these constraints by the following static analyses on seed programs:

\begin{itemize}
    \item $\mathrm{ArrayLength}(A, n)$: Check for statements that indicate $n$ as the size of $A$, such as \func{malloc}, \func{sizeof}, and \func{strlen}.
    \item $\mathrm{ArrayIndex}(A, i)$: If $i$ is a scalar and if its value is always smaller than the length of $A$.
    \item $\mathrm{FileName}(S)$: If $S$ is assigned the string \texttt{input\_file} or \\ \texttt{output\_file} (\autoref{fig: prompt}).
    \item $\mathrm{FormatString}(S)$: If $S$ contains the character \texttt{\%}.
    \item $\mathrm{AllocSize}(n)$: \tool infers this constraint by varying the arguments of scalar types between their minimum and maximum values, and then executing them with the fuzzing corpus and observing memory allocation sizes. If the sizes differ significantly, \tool infers this constraint.
    \item $\mathrm{FileDesc}(fd)$: Examine the data flow of the return value of the calls that are related to file descriptors, such as \func{open} and \func{fileno}.
\end{itemize}
If \tool infers multiple constraints on the same argument, it takes the constraint that has been inferred the most times.

\subsubsection{Constrained Argument Conversion} \label{sec:conversion}
\tool converts the arguments of constant values into variables of the same corresponding type where the variables can accept arbitrary bytes from the fuzzers.
\tool implements a custom FuzzedDataProvider~\cite{fdp} to segment bytes from the fuzzers into multiple sections and convert each section into a value of its designated type. To generate values of variables with statically known sizes, such as scalars and fixed-size arrays, it consumes bytes of the required size and statically casts them.
For dynamically sized variables, it consumes bytes until it encounters specific magic bytes.
After that, \tool adjusts the value to satisfy the inferred constraint on the argument. For arguments with the FileName, FormatString, AllocSize, or FileDesc constraints, \tool retains their original constant values. For arguments with the ArrayLength or ArrayIndex constraints, \tool ensures that each value is no larger than the corresponding array length.
For each converted argument, \tool attempts to provide them with several different random values.
If \tool's sanitizers detect an error, indicating an erroneous conversion, \tool cancels the conversion.

\subsubsection{Fuzzer Scheduling}\label{sec:fusion}
\tool consolidates the seed programs into a fuzz driver that schedules each seed program based on several specific bytes provided by the fuzzers.
To ensure efficient fuzzing, the fuzzing corpus in \autoref{sec: fuzzing_validation} serves as the initial input for this fuzz driver. \tool gathers the constant values of the converted arguments to form their initial corpora.
\begin{table*}[ht]
\caption{Overall results for \tool-generated fuzz drivers}\label{tab:overall-results}
\scriptsize
\begin{threeparttable}
\centering

\begin{tabular}{lllll|lll|lll|lll}
\toprule[1.5pt]
\multicolumn{5}{c|}{\textbf{Tested Library}}         & \multicolumn{3}{c|}{\textbf{Generated Programs}} & \multicolumn{3}{c|}{\textbf{Branch Coverage Comparison}} & \multicolumn{3}{c}{\textbf{Detected Bugs}} \\
\midrule[0.4pt]
\textbf{Name} & \textbf{Version} & \textbf{LoC} & \textbf{\#APIs} & \textbf{\#Branches} & \textbf{Total}        & \textbf{\#Seeds}       & \textbf{Query Cost / Time}   & \textbf{PromptFuzz}       & \textbf{OSS-Fuzz}      & \textbf{Hopper}      & \textbf{UC}         & \textbf{VB}        & \textbf{C}        \\
\midrule[0.8pt]
\textbf{curl}     &  8.4.0       &  154K   &    93    &     \num{26644}       &    \num{2550}          &   664(106)           &   \$3.58 / 12h32m    &   \textbf{\num{5283}(19.82\%)}          &  $\#20$ / 822(3.09\%)         &    \num{3383}(12.69\%)         &           0 &   0        &  0       \\
\textbf{libTIFF}     &  4.6.0       & 108K    &   195     &    \num{14204}        &  \num{2510}        &   153(71)            &   \$4.02 / 11h10m   &   \textbf{\num{7448}(52.43\%)}          & $\#1$ / \num{5740}(40.60\%)           & \num{3932}(27.68\%)            &       6     &     6      &   6      \\
\textbf{\scriptsize{libjpeg-turbo}}     &  3.0.1  & 144K    &  77      &    \num{10972}        &  \num{2730}        &   180(82)            &   \$4.98 / 22h44m   &   \num{5186}(47.26\%)          &  $\#9$ / \textbf{\num{6187}(56.39\%)}           &  \num{3971}(36.19\%)           &     4       &     2      &     2    \\
\textbf{sqlite3}     &   3.43.2      & 413K   &   289     &    \num{38056}        &  \num{2210}        &   404(74)            &     \$2.90 / 28h10m   &   \textbf{\num{28016}(73.61\%)}         &  $\#1$ / \num{9760}(25.64\%)          &  \num{10855}(28.52\%)           &      5      &     3      &    3     \\
\textbf{libpcap}     & 1.10.4       &  58K   &  84      &      \num{7816}         &  \num{2580}        &   151(49)            &    \$3.68 / 9h30m   &  \num{2974}(39.25\%)           &  $\#3$ / \num{3145}(41.51\%)           &  \textbf{\num{3686}(47.15\%)}           &     5       &   3        &   3      \\
\textbf{cJSON}     &   1.7.16      &  10K   &   76     &       \num{1020}         &  \num{2680}        &   209(54)            &    \$3.40 / 5h1m   &  846(82.94\%)             &  $\#1$ / 475(46.57\%)           &   \textbf{900(88.23\%)}         &         5   &    3       &   3      \\
\textbf{libaom}     &   3.7.0      &  530K   &  47      &      \num{61702}        &  \num{2290}        &   237(59)            &   \$4.11 / 16h26m    &  \textbf{\num{15811}(25.62\%)}          &  $\#1$ /  \num{10984}(18.01\%)         &  \num{7493}(12.14\%)           &     3       &    3       &   3      \\
\textbf{libvpx}     &  1.13.1       &  362K   &  40      &     \num{35544}        &   \num{3430}       &   396(98)            &    \$6.16 / 14h34m   &  \textbf{\num{7434}(20.91\%)}            &  $\#2$ / \num{4721}(13.32\%)           &  \num{3603}(10.13\%)           &    4        &     4      &     4    \\
\textbf{c-ares}     &  1.20.0       &  59K   &  61      &      \num{4038}         &   \num{1590}       &   126(38)            &   \$2.47 / 8h42m  &  \num{2141}(53.02\%)            &  $\#2$ / 791(22.80\%)           &  \textbf{\num{2932}(72.61\%)}           &         3   &      2     &  2       \\
\textbf{zlib}     &   1.3      &   30K  &    87    &        \num{2894}    &        \num{1630}          &   259(82)            &     \$2.41 / 11h10m   &  \num{2210}(76.36\%)             & $\#9$ /  \num{1525}(52.80\%)         &   \textbf{\num{2284}(78.92\%)}          &      1      &     0      &    0     \\
\textbf{re2}     &   bc0faab      &  28K   &  70      &     \num{4940}       &     \num{2140}          &   101(23)            &     \$3.30 / 13h2m   &  \num{3192}(64.61\%)              &  $\#1$ / \textbf{\num{3900}(78.94\%)}         &   \num{3403}(68.88\%)          &      0      &      0     &   0      \\
\textbf{lcms}     &   2.15      &  45K   &   286     &      \num{8806}            &     \num{9170}     &   402(96)            &   \$14.04 / 34h42m   &  \textbf{\num{3742}(42.49\%)}              &  $\#8$ / \num{3049}(34.62\%)         &  \num{2043}(23.20\%)          &   2         &       0    &    0     \\
\textbf{libmagic}     &  FILE5\_45   &  33K   &   18     &   \num{7440}         &   \num{2010}         &   217(32)            &   \$2.41 / 6h53m   &   \textbf{\num{4697}(63.67\%)}               & $\#3$ / \num{4628}(62.74\%)         &  \num{4377}(58.83\%)           &     4       &       4    &    1     \\
\textbf{libpng}     &  1.6.40       &  57K   &   246     &   \num{7732}         &   \num{3560}         &   286(99)            &   \$5.68 / 9h32m   &   \textbf{\num{3906}(50.51\%)}              & $\#1$ / \num{1967}(25.44\%)           &   \num{3847}(49.75\%)          &       2     &        0   &    0     \\
\midrule[0.8pt]
\textbf{Total} & - & 2M & \num{1669} & \num{231808} & \num{41080} & \num{3785}(963) & \$63.14 / 204h8m & \textbf{\num{92886}(40.07\%)} & \num{57694}(24.88\%)  & \num{56709}(24.46\%) & \textbf{44} & \textbf{30}  & \textbf{27} \\
\bottomrule[1.5pt]
\end{tabular}
\begin{tablenotes}
\item
\textbf{Seeds} = The number of seed programs present in the seed bank (and the count of \textit{unique seeds} among them);
\textbf{UC} = Number of reported unique crashes; \textbf{VB} = Number of valid bugs identified by manually review; \textbf{C} = Confirmed bugs after reported to the corresponding communities; The number of fuzzer drivers crafted for each library in \textsf{OSS-Fuzz} is prefixed with the '\#' symbol.
\end{tablenotes}

\end{threeparttable}
\end{table*}

\section{Implementation}
We implemented \tool in \num{17595} lines of Rust code and have made the source code available in our repository ~\cite{promptfuzz}.
The following sections will introduce some essential components implemented in \tool.

\subsection{AST Visitor}
\tool parses the Abstract Syntax Tree (AST) of program code and utilizes the \textsf{clang\_ast} crate~\cite{clang_ast} to deserialize the AST.
Once the AST is deserialized in Rust,
we implement an AST visitor to traverse the code's ASTs and extract node attributes.
This AST visitor enables \tool to achieve argument constraint inference as discussed in \autoref{sec:infer}.
Additionally, \tool performs source code transformation,
    as discussed in \autoref{sec:fusion},
    by utilizing the source code locations embedded in the attributes of AST nodes.
Building upon the AST visitor,
    we construct Control Flow Graphs (CFGs) for the programs and employ an intra-procedural data flow analysis engine on the CFG.
The CFG allows us to analyze the critical path,
    while the data flow analysis engine assists in analyzing the dependency between library API calls.

\subsection{File Sanitizer}\label{sec:fsan}
In addition to the use of ASan and UBSan during the sanitization process of \tool,
    we have also implemented a File-Sanitizer (\textsf{FSan}) to identify instances of error file operations, such as file descriptor leaks.
These errors are often responsible for performance degradation but are not detectable by ASan or UBSan.
Given that these errors significantly impact the effectiveness of the fuzz drivers generated by \tool, we have implemented FSan to identify these issues.
FSan achieves this by tracking the data flows of file descriptors, file streams, and file names, and by instrumenting detection code at the end of their lifespan within the source code.

\section{Evaluation}
In this section,
we conducted comprehensive evaluations to demonstrate the effectiveness of \tool.
Firstly,
    we evaluated \tool on 14 widely-used open-source libraries that have undergone extensive fuzzing through OSS-Fuzz~\cite{oss_fuzz} over several years.
We compared the code coverage achieved by \tool's fuzz drivers with other approaches for fuzz driver generation.
Secondly,
    we evaluated the ability of the fuzz drivers generated by \tool to find bugs.
Lastly,
    we analyzed the key components of \tool to demonstrate how each component contributes to its overall effectiveness.
    
All experiments were conducted on a server with 48-core CPUs clocked at 2.50GHz and 128 GB of RAM, running the 64-bit version of Ubuntu 20.04 LTS. \textsf{LibFuzzer}~\cite{libfuzzer} was the grey-box fuzz engine used in all evaluations.

\subsection{Overall Results}

We configured the \textsf{gpt-3.5-turbo-0613} and \textsf{gpt-3.5-turbo-16k-0613} models as the LLMs used for program generation.
When a query's tokens are shorter than the length limit of \textsf{gpt-3.5-turbo-0613},
    we chose \textsf{gpt-3.5-turbo-0613}; otherwise, we chose \textsf{gpt-3.5-turbo-16k-0613}, which comes at a higher cost but allows for a longer length limit.
We set the temperature parameter at 0.9 for the LLMs and sampled 10 programs per query.
The default API combination length was 5, the exponent $E$ in the power schedule was 1, and the initial prompt was constructed using a combination of 5 randomly selected library API functions.
In the evaluations of \tool,
    the fuzzing loop was operated continuously until 10 consecutive iterations passed without discovering new branches, without any restrictions on time and query budget.
Each library's consolidated fuzz driver was executed under a 24-hour timeout.
Every experiment was repeated five times to mitigate statistical errors, and the average results were reported.

Under the experimental setup,
    we used \tool to generate fuzz drivers for 14 open-source libraries and detect bugs.
The results of these experiments are summarized in \autoref{tab:overall-results},
    which provides the statistics about the tested libraries, the generated fuzz drivers, branch coverage, and bug detection results.
In total, \tool successfully \textbf{generated \num{3785} seed programs for these 14 libraries within 204 hours with the cost of \$63.14 for querying the LLMs} (\$4.15 per library on average)
\footnote{At the time of experiments, the input and output prices for gpt-3.5-turbo-0613 were 0.0015 and 0.002 per thousand tokens respectively, and the prices for gpt-3.5-turbo-16k-0613 are 0.003 and 0.004 per thousand tokens respectively.}.
Overall,          
    the fuzz drivers generated by \textbf{\tool achieved a branch coverage of 40.07\% on the tested libraries},
    which was 1.61x greater than \textsf{OSS-Fuzz} and 1.63x greater than \textsf{Hopper},
    and detected 30 previously unknown bugs during 24-hour experiments.
All bugs found have been reported to the corresponding communities.
In the following sections,
    we will detail the results of our evaluations.

\begin{table*}[!htb]
\caption{Previous known bugs found by \tool}
\footnotesize

\begin{threeparttable}
\begin{tabular}{lllllll}
\toprule[1.5pt]
\textbf{ID}  & \textbf{Library}  &\textbf{Location} & \textbf{Buggy Function}                      & \textbf{Bug Type}        & \textbf{Status} & \textbf{Track ID}
\\
\midrule[0.8pt]
1.  & libaom        &aom\_dsp/x86/highbd\_varianc\_sse2.c:49:7 & highbd\_8\_variance\_sse2                & Segment Violation                & Confirmed  & \href{https://bugs.chromium.org/p/aomedia/issues/detail?id=3489}{[3489]}                                \\
2.  & libaom        &av1/encoder/ratectrl.c:2501:7           & av1\_rc\_update\_framerate               & Uninitialized Stack & Confirmed  & \href{https://bugs.chromium.org/p/aomedia/issues/detail?id=3509}{[3509]}                                \\
3.  & libaom        &av1/encoder/encoder.h:3886:12           & timebase\_units\_to\_ticks               & Integer Overflow    & Confirmed  & \href{https://bugs.chromium.org/p/aomedia/issues/detail?id=3510}{[3510]}                                \\
4.  & libvpx        &vp8/vp8\_dx\_iface.c:133:3                & vp8\_peek\_si\_internal                  & Segment Violation                & Confirmed  & \href{https://bugs.chromium.org/p/webm/issues/detail?id=1817}{[1817]}                                   \\
5.  & libvpx        &vp8/vp8\_dx\_iface.c:252:47               & update\_fragments                        & Buffer Overflow     & Confirmed  & \href{https://bugs.chromium.org/p/webm/issues/detail?id=1827}{[1827]}                                   \\
6.  & libvpx        &vp8/vp8\_cx\_iface.c:951:56               & vp8e\_encode                             & Integer Overflow    & Confirmed  & \href{https://bugs.chromium.org/p/webm/issues/detail?id=1828}{[1828]}                                   \\
7.  & libvpx        &vp8/encoder/encodeframe.c:448:18        & encode\_mb\_row                          & Integer Overflow    & Confirmed  & \href{https://bugs.chromium.org/p/webm/issues/detail?id=1831}{[1831]}                                   \\
8. & libmagic      &apprentice.c:3289:11                    & apprentice\_map                          & Buffer Overflow     & Waiting    & \href{https://bugs.astron.com/view.php?id=481}{[481]}                                                   \\
9. & libmagic      &magic.c:617:19                          & magic\_setparam                          & Buffer Overflow     & Waiting    & \href{https://bugs.astron.com/view.php?id=482}{[482]}                                                   \\
10. & libmagic      &apprentice.c:3358:6                     & check\_buffer                            & Buffer Overflow     & Confirmed  & \href{https://bugs.astron.com/view.php?id=483}{[483]}                                                   \\
11. & libmagic      &softmagic.c:1675:11                     & mget                                     & Integer Overflow    & Waiting    & \href{https://bugs.astron.com/view.php?id=486}{[486]}                                                   \\
12. & libTIFF       &tif\_unix.c:222:12                       & TIFFOpen                                 & Out of Memory                 & Confirmed  & \href{https://gitlab.com/libtiff/libtiff/-/issues/614}{[614]}                                           \\
13. & libTIFF       &tif\_pixarlog.c:776:28                   & PixarLogSetupDecode                      & Out of Memory                 & Confirmed  & \href{https://gitlab.com/libtiff/libtiff/-/issues/619}{[619]}                                           \\
14. & libTIFF       &tif\_read.c:546:10                       & TIFFReadEncodedStrip                     & Out of Memory                 & Confirmed  & \href{https://gitlab.com/libtiff/libtiff/-/issues/620}{[620]}                                           \\
15. & libTIFF       &tif\_getimage.c:621:14                & TIFFReadRGBAImageOriented                & Out of Memory                 & Confirmed  & \href{https://gitlab.com/libtiff/libtiff/-/issues/620}{[620]}                                           \\
16. & libTIFF       &ti\_strip.c:333:9                       & TIFFRasterScanlineSize64               & Out of Memory                 & Confirmed  & \href{https://gitlab.com/libtiff/libtiff/-/issues/621}{[621]}                                           \\
17. & libTIFF       &tif\_getimage.c:3345:9                   & TIFFReadRGBATileExt                      & Segment Violation                & Confirmed  & \href{https://gitlab.com/libtiff/libtiff/-/issues/622}{[622]}                                           \\
28. & sqlite3       &sqlite3.c:178513:23                     & sqlite3\_unlock\_notify                  & Null Pointer Dereference  & Confirmed  & \href{https://www.sqlite.org/forum/forumpost/e77a5c3445}{[e77a5]}                                       \\
19. & sqlite3       &sqlite3.c:133068:23                     & sqlite3\_enable\_load\_extension         & Null Pointer Dereference  & Confirmed  & \href{https://www.sqlite.org/forum/forumpost/9ce835fe96}{[9ce83]} \\
20. & sqlite3       &sqlite3.c:174382:23                     & sqlite3\_db\_config                      & Null Pointer Dereference  & Confirmed  & \href{https://www.sqlite.org/forum/forumpost/5e3fc453a6}{[5e3fc]} \\
21. & c-ares        &lib/ares\_getaddrinfo.c:2173:8           & config\_sortlist                         & Memory Leak         & Confirmed  & \href{https://github.com/c-ares/c-ares/commit/d62627e8b39ef793c3b1c7b054724b0d581eb4fb}{[d62627]} \\
22. & c-ares        &lib/ares\_getaddrinfo.c:2184:8           & config\_sortlist                         & Memory Leak         & Confirmed  & \href{https://github.com/c-ares/c-ares/commit/d62627e8b39ef793c3b1c7b054724b0d581eb4fb}{[d62627]} \\
23. & libjpeg-turbo &turbojpeg.c:2245:46                     & tj3DecodeYUV8                            & Integer Overflow    & Confirmed  & \href{https://github.com/libjpeg-turbo/libjpeg-turbo/security/advisories/GHSA-x7cp-qgf3-9896}{[78eaf0]} \\
24. & libjepg-turbo &turbojpeg-mp.c:366:29                   & tj3LoadImage16                           & Out of Memory                 & Confirmed  & \href{https://github.com/libjpeg-turbo/libjpeg-turbo/issues/735}{[735]} \\
25. & libpcap       &pcap-linux.c:381:32                     & pcap\_create                             & File Descriptor Leak           & Confirmed  & \href{https://github.com/the-tcpdump-group/libpcap/issues/1233}{[1233]} \\
26. & libpcap       &pcap-linux.c:354:6                      & pcapint\_create\_interface               & Null Pointer Dereference  & Confirmed  & \href{https://github.com/the-tcpdump-group/libpcap/issues/1239}{[1239]} \\
27. & libpcap       &pcap-util.c:466:60                      & pcapint\_fixup\_pcap\_pkthdr             & Misaligned Address  & Confirmed  & - \\
28. & cJSON         &cJSON.c:394:28                          & cJSON\_SetNumberHelper                   & Type Error Cast          & Confirmed  & \href{https://github.com/DaveGamble/cJSON/issues/805}{[805]} \\
29. & cJSON         &cJSON.c:2448:30                         & cJSON\_CreateNumber                      & Type Error Cast          & Confirmed  & \href{https://github.com/DaveGamble/cJSON/issues/806}{[806]} \\
30. & cJSON         &cJSON.c:1892:83                         & \scriptsize{cJSON\_DeleteItemFromObjectCaseSensitive} & TimeOut             & Confirmed  & \href{https://github.com/DaveGamble/cJSON/issues/807}{[807]} \\
31*. & libaom        &av1/encoder/encoder.c:2605:16           &encode\_without\_recode                   & Segment Violation        & Confirmed & \href{https://bugs.chromium.org/p/aomedia/issues/detail?id=3534}{[3534]} \\
32*. & libvpx        &vpx\_tpl.c:140:29                        &vpx\_free\_tpl\_gop\_stats                    & Segment Violation        & Confirmed & \href{https://bugs.chromium.org/p/webm/issues/detail?id=1837}{[1837]} \\
33*. & curl          &urlapi.c:1245:3                         &parseurl                                  & Assertion Failure              & Confirmed & \href{https://github.com/curl/curl/pull/12775}{[12775]} \\
\bottomrule[1.5pt]
\end{tabular}
\begin{tablenotes}
\item
\textbf{Location}:  The source file location of the crash point. \textbf{Track ID}: The ID used for tracking this issue in their corresponding bug tracker system.
\item 
The bugs with IDs 31, 32 and 33 were detected through 15 days of fuzzing.
\end{tablenotes}

\end{threeparttable}
\label{tab:bugs}
\end{table*}
\subsection{Effectiveness on Code Coverage}
To evaluate how effective the fuzz drivers generated by \tool on code coverage are,
    we compared the branch coverage of libraries against the manually crafted fuzzers in \textsf{OSS-Fuzz} and the state-of-the-art automatic library fuzzing solution: \textsf{Hopper}.
During the evaluation,
    we ran fuzz drivers of \textsf{OSS-Fuzz} on each library for the same 24-hour period.
If there are multiple fuzz drivers for a library in \textsf{OSS-Fuzz},
    we ensured each driver ran independently on a distinct CPU core for the same 24-hour duration.
For \textsf{Hopper}, which assembles fuzz drivers and performs fuzzing on libraries simultaneously,
     we ran \textsf{Hopper} on each library for a period of 24 hours plus the time \tool took to generate fuzz drivers for that library, thereby ensuring a fair comparison.

The evaluation results are shown in \autoref{tab:overall-results}.
In comparing \tool against fuzz drivers from \textsf{OSS-Fuzz} and Hopper on 14 libraries, \tool demonstrates the highest branch coverage in 8 out of the 14 libraries. Among the remaining 6 libraries where \tool did not top the list, the coverage shortfall in \file{cJSON} and \file{zlib} was marginal. For \file{libjpeg-turbo}, \file{libpcap}, \file{re2}, and \file{c-ares}, the coverage gap was within a range of 1000 code branches, a margin that is considered acceptable within the scope of this study.

Compared to \textsf{OSS-Fuzz},
    \tool achieved higher branch coverage (40.07\%) than \textsf{OSS-Fuzz} (24.88\%) in the libraries totally.
    The results become even more remarkable given the fact that multiple fuzz drivers built for \file{curl}, \file{zlib}, and \file{lcms} are provided in OSS-Fuzz and these libraries thus have been fuzzed for more than 24 hours.
This higher branch coverage achieved by \tool can be primarily attributed to its capability of generating programs that cover a wide range of library usage scenarios.

Compared to \textsf{Hopper},
    which automatically synthesizes fuzz drivers through interpretative fuzzing,
    \tool achieved higher total branch coverage (40.07\% vs.24.46\%) as well.
There are two main reasons for \tool's better performance compared to Hopper.
Firstly,
    leveraging internal knowledge of LLMs,
    \tool effectively extracts complex information about API interdependency in various library API functions such as \file{libTIFF}, \file{sqlite3}, and \file{lcms},
    whereas \textsf{Hopper} blindly infers them.
Secondly, \textsf{Hopper} was unable to generate the necessary code pattern for libraries that require iterative calls of library API functions, such as \file{libaom}, \file{libvpx}, and \file{re2}, because it lacks support for conditional grammars. 
In contrast, \tool can support all types of control flow transitions.
Overall,
    \tool generates fuzz drivers with higher overall code coverage than \textsf{OSS-Fuzz} and \textsf{Hopper}.

\subsection{Effectiveness on Bug Detection}
To demonstrate the effectiveness of \tool in detecting library bugs, we analyzed the crashes reported by the fuzz drivers generated by \tool during their fuzzing time. Duplicate crashes were removed by examining the call traces. Throughout the 24-hour fuzzing period, 44 unique crashes were reported. After manually reviewing the code and documentation to verify the validity of these unique crashes, we identified 30 of them as valid bugs. All identified bugs have been reported to the respective communities for confirmation and resolution. At the time of writing, 27 identified bugs have been confirmed, while the remaining three await responses. The details of these bugs are available in \autoref{tab:bugs}.

\subsubsection{False Positive Analysis}
We analyzed the root causes of the ineffective warnings produced by previous fuzzing.
Among the 14 ineffective warnings,
    there were 8 warnings resulting from dereferencing null pointers returned by library API calls.
As demonstrated in \autoref{sec:eval_infer},
    the conversion of arguments of library API calls significantly enhances the bug-finding capability of the fuzz drivers,
    but it also increases the likelihood of library API calls entering error states and returning null pointers.
If the subsequent library API calls access these null pointers without implementing handling for those null pointer arguments, spurious crashes may occur.
We must note that we do not consider these crashes false positives in \tool's bug detection.
Instead,
    they are robustness issues stemming from the library API functions failing to handle the passed null pointers.
Excluding the 8 warnings reported due to library API robustness issues,
    \textbf{only 6 crashes were identified as false positives in \tool's bug detection. 
    We argue that \tool achieves a detection accuracy of 86.36\% (38/44)}.
Among these 6 false positives,
    2 crashes were caused by constraints that \tool failed to infer from the libraries.
Those constraints haven't been deduced because LLMs failed to generate the correct usage for the corresponding library API functions,
    hence causing the conversion of their arguments to trigger violations.
The remaining 4 positives were considered misuse of the target library that escaped \tool's validation due to their complex triggering mechanisms.
For examples,
    a false positive found in \textsf{zlib} could only be triggered by special values\footnote{https://github.com/madler/zlib/issues/904} and an issue found in \textsf{libpng} requires passing a consistent set of arguments to \func{png\_write\_png}\footnote{https://github.com/pnggroup/libpng/issues/491}.

\begin{table*}[!htb]
\caption{Impact of eliminated erroneous programs and inferred argument constraints}
\begin{threeparttable}
\footnotesize

\begin{tabular}{l|ccccc|cccccc}
\toprule[1.5pt]
\multirow{2}{*}{\textbf{Library}}  & \multirow{2}{*}{\textbf{Total}} & \multicolumn{3}{c}{\textbf{Eliminated Erroneous Programs}}  & \multirow{2}{*}{\textbf{Remain}}  & \multicolumn{6}{c}{\textbf{Inferred Argument Constraints}}                           \\ \cline{3-5} \cline{7-12}
                     &    & \textbf{Syntax}         & \textbf{Fuzzing}        & \textbf{Coverage}   &     & \textbf{ArrayLength}          & \textbf{ArrayIndex}        & \textbf{Format}     & \textbf{FileName}   & \textbf{FileDesc}  & \textbf{AllocSize}   \\
\midrule[0.8pt]
curl                     &\num{2550} & \num{1291}           & 472           & 123        &664     & 13 / 13 / 0 & 0 / 0 / 0 & 10 / 5 / 0 & 2 / 2 / 1  & 3 / 3 / 0 & 0 / 0 / 0   \\
libTIFF                  &\num{2510}  & \num{1303}           & 994           & 60        &153      & 25 / 19 / 1 & 0 / 0 / 0 & 0 / 0 / 0  & 6 / 3 / 2  & 3 / 1 / 1 & 4 / 4 / 0   \\
libjpeg-turbo            &\num{2730}  & \num{1948}           & 267           & 335       &180      & 25 / 23 / 3 & 0 / 0 / 2 & 0 / 0 / 0  & 8 / 8 / 0  & 0 / 0 / 0 & 12 / 10 / 0 \\
sqlite3                  &\num{2210}  & 920            & 638           & 248       &404      & 25 / 10 / 0 & 0 / 0 / 0 & 7 / 3 / 0  & 12 / 2 / 0 & 0 / 0 / 0 & 4 / 0 / 0   \\
libpcap                  &\num{2580}  & \num{1232}           & 583           & 614       &151      & 3 / 3 / 0   & 0 / 0 / 0 & 0 / 0 / 0  & 5 / 5 / 1  & 0 / 0 / 2 & 0 / 0 / 0   \\
cJSON                    &\num{2680}  & 562            & \num{1630}          & 279       &209      & 7 / 7 / 0   & 0 / 0 / 0 & 0 / 0 / 0  & 0 / 0 / 0  & 0 / 0 / 0 & 1 / 1 / 0   \\
libaom                   &\num{2290}  & \num{1437}           & 244           & 372       &237      & 7 / 7 / 0   & 1 / 1 / 0 & 0 / 0 / 0  & 0 / 0 / 1  & 0 / 0 / 0 & 0 / 0 / 0   \\
libvpx                   &\num{3430}  & \num{1943}           & 676           & 415       &396      & 3 / 3 / 0   & 1 / 1 / 0 & 0 / 0 / 0  & 0 / 0 / 0  & 0 / 0 / 0 & 0 / 0 / 0   \\
c-ares                   &\num{1590}  & 863            & 541           & 60        &126      & 21 / 20 / 0 & 0 / 0 / 1 & 0 / 0 / 0  & 0 / 0 / 0  & 2 / 2 / 0 & 0 / 0 / 0   \\
zlib                     &\num{1,630}   & 709            & 477           & 185      &259       & 26 / 25 / 0 & 0 / 0 / 1 & 2 / 1 / 0  & 2 / 2 / 1  & 1 / 1 / 0 & 2 / 2 / 0   \\
re2                      &\num{2140}   & \num{1182}           & 814           & 43       &101       & 6 / 5 / 0   & 0 / 0 / 0 & 0 / 0 / 0  & 0 / 0 / 0  & 0 / 0 / 2 & 0 / 0 / 0   \\
lcms                     &\num{9170}   & \num{6627}           & \num{2048}          & 93       &402       & 25 / 23 / 2 & 3 / 3 / 3 & 1 / 1 / 0  & 6 / 6 / 0  & 0 / 0 / 0 & 7 / 4 / 0   \\
libmagic                 &\num{2010}   & \num{1295}           & 276           & 222      &217       & 2 / 2 / 0   & 0 / 0 / 0 & 0 / 0 / 0  & 6 / 6 / 0  & 1 / 1 / 0 & 0 / 0 / 0   \\
libpng                   &\num{3560}   & \num{2521}           & 600           & 153      &286       & 16 / 7 / 0  & 0 / 0 / 1 & 0 / 0 / 0  & 1 / 1 / 0  & 0 / 0 / 0 & 4 / 4 / 0   \\
\midrule[0.8pt]
\textbf{Total}         & \textbf{\num{41080}} & \textbf{\num{23833}}          & \textbf{\num{10260}}        & \textbf{\num{3202}}   &\textbf{\num{3785}} & \textbf{197 / 167 / 5} & \textbf{5 / 5 / 8} & \textbf{20 / 10 / 0} & \textbf{48 / 35 / 6} & \textbf{10 / 8 / 5} & \textbf{34 / 25 / 0} \\   
\bottomrule[1.5pt]
\end{tabular}
\begin{tablenotes}
\item
The column of \textbf{total} refers to the number of generated programs. The column of \textbf{remain} refers to the number of programs that passed our validation.
The numbers separated by slashes represent the counts of \textbf{the ground truth constraints}, \textbf{\tool correctly inferred constraints}, and \textbf{\tool error-inferred constraints}, respectively.
\end{tablenotes}
\end{threeparttable}
`

\label{tab:san_and_infer}
\end{table*}
\subsubsection{False Negative Analysis}
In our preceding fuzzing experiments, OSS-Fuzz found no bugs, while Hopper identified 5 valid bugs.
To investigate the false negative rate of \tool in detecting library bugs,
    we selected the 5 bugs identified by Hopper and the 17 bugs presented in \textsf{Hopper}'s paper, all of which were confirmed by developers, as our evaluation benchmark.
Out of the five bugs found by \textsf{Hopper} in our experiments, \tool detected three bugs\footnote{IDs 17, 19, and 30 as shown in \autoref{tab:bugs}}, but missed two.
Our manual inspection of the two missed bugs revealed that \tool had successfully generated the code containing the associated bug patterns, but one was eliminated by \tool's fuzzing validation, while the other did not generate the input required to trigger a crash.
The 17 bugs presented in \textsf{Hopper}'s paper were attached with commit IDs related to their bug reporting, and the details of those bugs can be obtained via their commit messages.
For each one of them,
    we determined that \tool could detect it if its bug pattern appeared in the fuzz drivers generated by \tool.
Consequently, \tool was able to detect 15 out of these 17 genuine library bugs.
The two false negatives were caused by LLM's failure to generate the related buggy code pattern.
For instance,
    in the \textsf{lcms} with 286 library API functions,
    the API \func{cmsStageAllocMatrix} hasn't been generated with related buggy code patterns\footnote{https://github.com/mm2/Little-CMS/issues/354} due to the huge API combination search space.
In total, \tool was able to detect 18 of 22 bugs found by the state-of-the-art library fuzzing approaches.

\subsubsection{Long-term Fuzzing}
 To demonstrate the consistent detection of new library bugs with longer fuzzing durations, we conducted an additional 15-day fuzzing session on \tool's fuzz drivers. Consequently, \tool reported five more unique crashes, three of which were identified as valid library bugs. These three new bugs were also confirmed by developers, and the details of them can be obtained in \autoref{tab:bugs}.

\subsection{Effectiveness of \tool's components}
In this section,
    we conducted experiments to investigate the impact of the proposed techniques on the effectiveness of \tool.
\autoref{tab:san_and_infer} presents the detailed analysis results for the eliminated erroneous programs and the inferred argument constraints of \tool in previous experiments.

\begin{figure*}[t]
  \centering \includegraphics[width=1\linewidth]{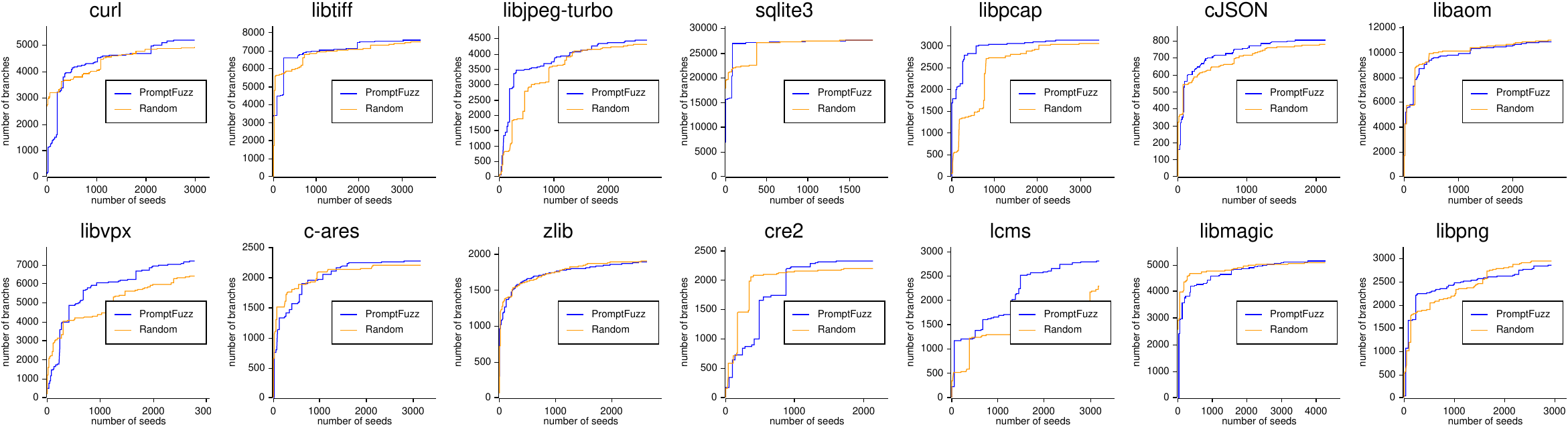}
  \caption[power schedule experiment]{Branches covered by seed programs generated by \tool under coverage-guided mutation and blind mutation}
  \label{fig:mutation_cmp}
\end{figure*}

\subsubsection{Erroneous Programs Elimination}\label{sec:eval_eliminate}
The numbers of programs eliminated by each process of \tool's validation are shown in \autoref{tab:san_and_infer}.
It can be observed that the majority of the erroneous programs (\num{23833}, 63.90\%) were eliminated due to syntax errors.
Additionally,
    the \textit{fuzzing} validation process of \tool identified \num{10260} programs (27.51\%) that exhibited abnormal runtime behaviors.
Furthermore,
    \num{3202} programs were eliminated by the \textit{coverage} validation due to insufficient code coverage.
Among the \num{10260} programs eliminated by \tool's \textit{fuzzing} validation,
    we analyzed their crash reports to investigate the factors contributing to the abnormal runtime behaviors.
The most prevalent issues detected were segmentation violations (\num{3394}, 33.07\%) and memory leaks (\num{3003}, 29.26\%) identified by the sanitizers.
Specifically, \tool's FSan (detailed in \autoref{sec:fsan}) detected 324 programs containing opened file leaks.

To investigate whether the programs were correctly eliminated,
    we conducted a study in which we randomly selected 10 programs for each library eliminated by the \textit{fuzzing} validation and 10 programs eliminated by the \textit{coverage} validation.
We reviewed the code of these programs and conducted careful debugging to determine if they had been properly eliminated.
The results revealed that almost all 140 programs eliminated by the \textit{fuzzing} validation contained misuses of library API functions.
The only exception is a latent resource leak detected by FSan in \file{libpcap}\footnote{https://github.com/the-tcpdump-group/libpcap/issues/1233}.
This genuine bug originates from file descriptor leaks resulting from the mismatched resource allocation and deallocation between the functions \func{pcap\_create} and \func{pcap\_close}.
Without FSan,
    such a hidden bug in the most commonly used code pattern in \textit{libpcap} would never have been uncovered.
For the 140 programs eliminated by the \textit{coverage} validation,
    108 of them were confirmed to have erroneous library usage and were correctly eliminated due to the presence of unreachable library API calls. 
Among these,
    25 were caused by incorrect library initialization,
    and 40 were caused by the wrong API context,
    and 43 were due to invalid library API configurations.
The remaining 32 programs were mistakenly eliminated because the fuzzer failed to generate input that can reach certain library API calls that are theoretically reachable within the time budget assigned for the \textit{fuzzing} validation process of \tool (see in \autoref{sec: fuzzing_validation}).

It is important to note that although the validation processes implemented in \tool may unintentionally exclude correctly functioning and genuine buggy programs, they significantly reduce spurious crashes during bug detection.

\subsubsection{Argument Constraint Inference}\label{sec:eval_infer}
In \autoref{sec:infer},
    we proposed the techniques to infer constraints imposed on arguments of library API functions,
    and convert the library API call arguments to receive random bytes from fuzzers.
To assess the accuracy of the \tool's constraint inference,
    we inspected documents of the tested libraries to collect the ground truth of API argument constraints.
As \autoref{tab:san_and_infer} shows,
    \tool achieves 91.24\% (250/274) precision and 79.61\% (250/314) recall on the inference of argument constraints.
The false positives are mainly owing to the absence of argument identifiers in the declarations of library API functions.
This deficiency hampers the ability of LLMs to comprehend the functionalities of these arguments, consequently leading to inaccurate library API usage generation.
Noticeably,
    constraints inferred by \tool is intended to limit the incorrect conversion of constants of API arguments.
    Therefore, false positives in inferred constraints will not cause additional spurious crashes.
The false negatives were primarily because LLMs have not generated code for the relevant library API functions yet, and they rarely resulted in false positives in bug detection.
Having those inferred constraints, \tool can convert the library API arguments to receive random bytes without violating the constraints imposed by developers.

To quantify the number of bugs identified through the argument conversion of library API calls,
    we examined the program code to determine which arguments were responsible for the crashes.
As a result,
    15 out of the 33 identified bugs could only be detected using the additional converted arguments.
For instance,
    all reported crashes in \file{libvpx} and \file{libaom} were triggered by incompatible flags and configurations passed to the codec.
Without converting these arguments, 
    the generated fuzz drivers would never have the chance to trigger them.
These results highlight that,
    while the programs generated and eliminated by \tool can serve as suitable targets for fuzzing,
    the ability of the resulting fuzz drivers to uncover new bugs is limited.
\tool's capability to convert additional arguments of library API functions to receive random bytes from fuzzers significantly contributed to the discovery of new bugs.

\subsubsection{Coverage Guide Prompt Mutation}
\tool develops a \textit{coverage-guided mutation} to instruct LLMs in generating valuable programs.
To evaluate its effectiveness,
    we experimented by comparing it with a random blind mutation approach.
In this experiment,
    The blind mutation approach was configured to randomly select library API functions with the same default combination length.
To ensure fairness,
    both the coverage-guided mutation setup and the blind mutation setup were assigned the same query budget (i.e., \$5) and were executed until the budget was exhausted.
Additionally,
    the temperature of the LLMs was set to 0.1 to reduce the randomness of the LLMs, and each experiment was repeated 5 times.

\autoref{fig:mutation_cmp} displays the accumulated covered branches attained by the generated seed programs during the fuzz loops of \tool when configured with two different mutation methods.
When giving the same query budget,
    the \textit{coverage-guided mutation} outperformed random blind mutation in 11 out of the 14 libraries, with the exceptions being \file{libaom}, \file{zlib}, and \file{libpng}.
Despite the low growth rate of branch coverage in the warm-up stages,
    \textit{coverage-guided mutation} surpassed random blind mutation in the 11 libraries due to the feedback obtained from the coverage and seed programs.
This enabled \tool to mutate prompts that incorporated meaningful combinations of API functions, creating programs that reached deeper library states.

The factors leading to the underperformance include the presence of loose coupling between API functions and the large number of API functions within these three libraries.
In \file{libaom},
    the API functions exhibit a high degree of coherence, and the interdependency between API functions is evident from their declarations. This clarity facilitates the generation of programs by LLMs, even when provided with randomly selected API functions.
For \file{libpng},
    the extensive number of API functions tend to trap the \textit{coverage-guided mutation} setup in local states, while random mutation allows exploration of a broader range of API functions.
This underperformance is expected to be resolved by allocating a larger query budget for LLMs.
Although the \textit{coverage-guided mutation} does not guarantee outperformance in all libraries, the experimental results demonstrate that it is the superior approach in most cases.

\section{Discussion \& Future Directions}
The prototype of \tool demonstrates the effectiveness of fuzz driver generation for open-source libraries. 
To give insights into more effective fuzzing approaches, we would like to discuss some essential aspects of \tool and highlight several possible future directions.

\noindent
\textbf{LLM models choice.} 
In our experiments, 
    we use \textsf{GPT3.5}~\cite{GPT3.5} as the LLM model to generate programs using prompts constructed by \tool.
    The performance of different models can vary significantly,
        leading to different outputs and results.
    More powerful LLM models are prone to bring a more profound understanding of prompted instructions and generate higher-quality programs,
        indicating that the performance of \tool may improve if more powerful LLM models are employed.

\noindent
\textbf{Detection accuracy enhancement.}
\tool achieves a high detection accuracy of 86.36\%, which is manageable for developers to mitigate library bugs. However, we have identified some strategies that could further enhance its accuracy. \textsf{Hopper} avoids invalid null pointer crashes by inserting assert statements after each library API call that returns a pointer. This approach can help \tool prevent ineffective warnings caused by null pointer dereferences. Several methods have been proposed for program repair using LLMs~\cite{xia2023automated, xia2023conversational, jin2023inferfix}. Drawing upon these ideas, we can help \tool further reduce spurious crashes.


\noindent
\textbf{Application scope extension.}
The evaluated libraries selected for our assessments are open-source.
The source code of these libraries is publicly available and was used as the training data for our chosen LLM model.
The evaluation results may differ if \tool is directly applied to closed-source libraries. Nevertheless, fine-tuning LLMs~\cite{finetune_intro, custom_finetune} could mitigate these issues.
Besides that,
    we found it is potential to apply \tool to more types of software, such as web applications or embedded systems.
\textsf{AFGen}~\cite{afgen} is a whole-function fuzzing approach that composes fuzz drivers for internal functions of applications.
Drawing from this idea, we are also able to generalize \tool to system or web applications by treating the internal functions as the interfaces.
\section{Related work}

\subsection{Fuzz Driver Generation.}
Several approaches have been proposed to facilitate the generation of fuzz drivers~\cite{fudge, fuzzgen, graphfuzz, intelligen, apicraft, winnie, utopia, hopper, afgen, titan_fuzz, oss-llm, rulf}. \textsf{Fudge}~\cite{fudge}, \textsf{FuzzGen}~\cite{fuzzgen}, and \textsf{UTopia}~\cite{utopia} generate fuzz drivers by extracting library usage from consumer code. For instance, \textsf{FuzzGen} constructs an Abstract API Dependence Graph ($A^2DG$) by analyzing the code of the Android Open Source Project (AOSP) and creates fuzz drivers by traversing the $A^2DG$. Meanwhile, \textsf{APICraft}~\cite{apicraft} and \textsf{Winnie}~\cite{winnie} utilize the library usage learned from execution traces to create fuzz drivers. Unfortunately, these approaches fail to consider libraries without external consumers. To overcome this limitation, some approaches~\cite{intelligen, graphfuzz, rulf} have been proposed that generate fuzz drivers without relying on external consumers. \textsf{GraphFuzz}~\cite{graphfuzz} relies on the library specification provided by users to compose fuzz drivers, while \textsf{RULF}~\cite{rulf} relies on strong type restraints in Rust to create fuzz drivers. However, these approaches require human integration or are limited to domain-specific libraries. In addition to the aforementioned approaches, \textsf{Hopper} synthesizes fuzz drivers by fuzzing an interpreter to compose valid library API calls. However, the vast search space of API functions and arguments limits the effectiveness of fuzz drivers generated by \textsf{Hopper}. Additionally, \textsf{TitanFuzz}~\cite{titan_fuzz} and the method proposed by Google~\cite{oss-llm} rely on LLMs to generate fuzz drivers, but they struggle to generate fuzz drivers uncover deep library bugs. Compared to these approaches, \tool generates fuzz drivers without requiring external consumers and domain-specific models while maintaining their effectiveness in bug detection.

\subsection{Deep Learning-based Software Testing.}
Deep learning techniques are increasingly being utilized in software testing.
\textsf{SparrowHawk}~\cite{SparrowHawk} and \textsf{Goshawk}~\cite{Goshawk} employs natural language processing (NLP) models to identify custom memory management functions within software projects, enhancing static code analysis. \textsf{CarpetFuzz}~\cite{carpetfuzz} utilizes NLP to extract API constraints from software documents and detects violations by analyzing the dependencies between API calls.
\textsf{Pythia}~\cite{atlidakis2020pythia} is a grammar-based REST API fuzzer that utilizes a seq2seq model to achieve grammar mutation.
In addition to these approaches that require training on specific deep learning models, several approaches are designed directly on pre-trained LLMs ~\cite{CodaMosa, xia2023universal, llm_parser, llm_edge}.
\textsf{Fuzz4All}~\cite{xia2023universal} and \textit{Joshua et.al} fuzz the program parser and language parser by utilizing LLMs to generate and mutate input to the parser.
\textsf{CodaMosa}~\cite{CodaMosa} employs LLMs to provide test cases for uncovered functions,
    addressing coverage plateaus caused by them.
\textsf{GPTFuzz}~\cite{llm_edge} tests the robustness of deep learning library API functions by using LLMs to generate vulnerable cases.
Compared with these approaches, \tool is a novel solution that aims for automatic fuzz driver generation.
It constructs a fuzz loop to iteratively generate fuzz drivers that cover a broader range of library code.
The fuzz drivers generated by \tool can effectively test various library usage while maintaining high bug detection accuracy.
\section{Conclusion}
This paper presents \tool, a coverage-guided fuzzer for automatic fuzz driver generation. PromptFuzz generates fuzz drivers through prompt fuzzing, a novel fuzz loop built upon LLMs. Guided by coverage feedback, \tool iteratively constructs prompts of LLMs to explore a wide range of API usage efficiently. We designed oracles for detecting erroneous programs generated by LLMs. By relying on the code synthesis capability of LLMs, \tool creates fuzz drivers without requiring consumer code or domain knowledge. The fuzz drivers generated by \tool achieve higher branch coverage, 1.61 times greater than that of OSS-Fuzz and 1.63 times greater than that of Hopper. Additionally, the fuzz drivers generated by \tool successfully detect 33 new genuine bugs out of 49 crashes, 30 out of which have been confirmed by their communities.

\bibliographystyle{acm}
\bibliography{main.bib}

\begin{thebibliography}{10}

\bibitem{llm_parser}
{\sc Ackerman, J., and Cybenko, G.}
\newblock Large language models for fuzzing parsers (registered report).
\newblock In {\em Proceedings of the 2nd International Fuzzing Workshop\/} (2023), FUZZING 2023.

\bibitem{amann2018systematic}
{\sc Amann, S., Nguyen, H.~A., Nadi, S., Nguyen, T.~N., and Mezini, M.}
\newblock A systematic evaluation of static api-misuse detectors.
\newblock {\em IEEE Transactions on Software Engineering 45}, 12 (2018), 1170--1188.

\bibitem{fuzzomatic}
{\sc Amiet, N.}
\newblock Introducing fuzzomatic: Using ai to automatically fuzz rust projects from scratch.
\newblock \url{https://research.kudelskisecurity.com/2023/12/07/introducing-fuzzomatic-using-ai-to-automatically-fuzz-rust-projects-from-scratch/}.

\bibitem{atlidakis2020pythia}
{\sc Atlidakis, V., Geambasu, R., Godefroid, P., Polishchuk, M., and Ray, B.}
\newblock Pythia: Grammar-based fuzzing of rest apis with coverage-guided feedback and learning-based mutations.
\newblock {\em arXiv preprint arXiv:2005.11498\/} (2020).

\bibitem{fudge}
{\sc Babi{\'c}, D., Bucur, S., Chen, Y., Ivan{\v{c}}i{\'c}, F., King, T., Kusano, M., Lemieux, C., Szekeres, L., and Wang, W.}
\newblock Fudge: fuzz driver generation at scale.
\newblock In {\em Proceedings of the 2019 27th ACM Joint Meeting on European Software Engineering Conference and Symposium on the Foundations of Software Engineering\/} (2019), pp.~975--985.

\bibitem{bi-etal-2019-incorporating}
{\sc Bi, B., Wu, C., Yan, M., Wang, W., Xia, J., and Li, C.}
\newblock Incorporating external knowledge into machine reading for generative question answering.
\newblock In {\em Proceedings of the 2019 Conference on Empirical Methods in Natural Language Processing and the 9th International Joint Conference on Natural Language Processing (EMNLP-IJCNLP)\/} (Nov. 2019), pp.~2521--2530.

\bibitem{aflfast}
{\sc B\"{o}hme, M., Pham, V.-T., and Roychoudhury, A.}
\newblock Coverage-based greybox fuzzing as markov chain.
\newblock In {\em Proceedings of the 2016 ACM SIGSAC Conference on Computer and Communications Security\/} (2016), p.~1032–1043.

\bibitem{gpt3}
{\sc Brown, T., Mann, B., Ryder, N., Subbiah, M., Kaplan, J.~D., Dhariwal, P., Neelakantan, A., Shyam, P., et~al.}
\newblock Language models are few-shot learners.
\newblock In {\em Advances in Neural Information Processing Systems\/} (2020), pp.~1877--1901.

\bibitem{oss2023}
{\sc Chang, O.}
\newblock Taking the next step: Oss-fuzz in 2023.
\newblock \url{https://security.googleblog.com/2023/02/taking-next-step-oss-fuzz-in-2023.html}.

\bibitem{contractBasedApr}
{\sc Chen, L., Pei, Y., and Furia, C.~A.}
\newblock Contract-based program repair without the contracts.
\newblock In {\em 2017 32nd IEEE/ACM International Conference on Automated Software Engineering (ASE)\/} (2017), pp.~637--647.

\bibitem{angora}
{\sc Chen, P., and Chen, H.}
\newblock Angora: efficient fuzzing by principled search.
\newblock In {\em IEEE Symposium on Security and Privacy (S\&P)\/} (San Francisco, CA, 5 2018).

\bibitem{Chen:2019:Matryoshka}
{\sc Chen, P., Liu, J., and Chen, H.}
\newblock {Matryoshka}: Fuzzing deeply nested branches.
\newblock In {\em ACM Conference on Computer and Communications Security (CCS)\/} (London, UK, 2019).

\bibitem{hopper}
{\sc Chen, P., Xie, Y., Lyu, Y., Wang, Y., and Chen, H.}
\newblock Hopper: Interpretative fuzzing for libraries.
\newblock In {\em ACM Conference on Computer and Communications Security (CCS)\/} (Copenhagen, Denmark, 2023).

\bibitem{custom_finetune}
{\sc Das, S.}
\newblock Fine tune large language model (llm) on a custom dataset.
\newblock \url{https://dassum.medium.com/fine-tune-large-language-model-llm-on-a-custom-dataset-with-qlora-fb60abdeba07}.

\bibitem{llm_edge}
{\sc Deng, Y., Xia, C., Yang, C., Zhang, S., Yang, S., and Zhang, L.}
\newblock Large language models are edge-case generators: Crafting unusual programs for fuzzing deep learning libraries.
\newblock In {\em 2024 IEEE/ACM 46th International Conference on Software Engineering (ICSE)\/} (2024).

\bibitem{titan_fuzz}
{\sc Deng, Y., Xia, C.~S., Peng, H., Yang, C., and Zhang, L.}
\newblock Large language models are zero-shot fuzzers: Fuzzing deep-learning libraries via large language models.
\newblock In {\em Proceedings of the 32nd ACM SIGSOFT International Symposium on Software Testing and Analysis\/} (2023), pp.~423--435.

\bibitem{oss-llm}
{\sc Dongge~Liu, J.~M., and Chang, O.}
\newblock Fuzz target generation using llms.
\newblock \url{https://google.github.io/oss-fuzz/research/llms/target_generation/}.

\bibitem{finetune_intro}
{\sc Ferrer, J.}
\newblock An introductory guide to fine-tuning llms.
\newblock \url{https://www.datacamp.com/tutorial/fine-tuning-large-language-models}.

\bibitem{bard}
{\sc Google}.
\newblock Google's bard.
\newblock \url{https://bard.google.com/}.

\bibitem{honggfuzz}
{\sc Google}.
\newblock Honggfuzz.
\newblock \url{https://github.com/google/honggfuzz}.

\bibitem{oss_corpus}
{\sc Google}.
\newblock How to prepare the seed corpus for oss-fuzz.
\newblock \url{https://google.github.io/oss-fuzz/getting-started/new-project-guide/#seed-corpus}.

\bibitem{good_fuzz_target}
{\sc Google}.
\newblock What makes a good fuzz target.
\newblock \url{https://github.com/google/fuzzing/blob/master/docs/good-fuzz-target.md}.

\bibitem{graphfuzz}
{\sc Green, H., and Avgerinos, T.}
\newblock Graphfuzz: Library api fuzzing with lifetime-aware dataflow graphs.
\newblock In {\em 2022 IEEE/ACM 44th International Conference on Software Engineering (ICSE)\/} (2022), pp.~1070--1081.

\bibitem{fuzzgen}
{\sc Ispoglou, K., Austin, D., Mohan, V., and Payer, M.}
\newblock {FuzzGen}: Automatic fuzzer generation.
\newblock In {\em 29th USENIX Security Symposium (USENIX Security 20)\/} (2020), pp.~2271--2287.

\bibitem{utopia}
{\sc Jeong, B., Jang, J., Yi, H., Moon, J., Kim, J., Jeon, I., Kim, T., Shim, W., and Hwang, Y.~H.}
\newblock Utopia: Automatic generation of fuzz driver using unit tests.
\newblock In {\em 2023 IEEE Symposium on Security and Privacy (SP)\/} (2022), IEEE Computer Society, pp.~746--762.

\bibitem{hallucination}
{\sc Ji, Z., Lee, N., Frieske, R., Yu, T., Su, D., Xu, Y., Ishii, E., Bang, Y.~J., Madotto, A., and Fung, P.}
\newblock Survey of hallucination in natural language generation.
\newblock {\em ACM Computing Surveys 55}, 12 (2023), 1--38.

\bibitem{jiang2018aprspace}
{\sc Jiang, J., Xiong, Y., Zhang, H., Gao, Q., and Chen, X.}
\newblock Shaping program repair space with existing patches and similar code.
\newblock In {\em Proceedings of the 27th ACM SIGSOFT International Symposium on Software Testing and Analysis\/} (New York, NY, USA, 2018), ISSTA 2018, Association for Computing Machinery, p.~298–309.

\bibitem{rulf}
{\sc Jiang, J., Xu, H., and Zhou, Y.}
\newblock Rulf: Rust library fuzzing via api dependency graph traversal.
\newblock In {\em 2021 36th IEEE/ACM International Conference on Automated Software Engineering (ASE)\/} (2021), IEEE, pp.~581--592.

\bibitem{jin2023inferfix}
{\sc Jin, M., Shahriar, S., Tufano, M., Shi, X., Lu, S., Sundaresan, N., and Svyatkovskiy, A.}
\newblock Inferfix: End-to-end program repair with llms.
\newblock {\em arXiv preprint arXiv:2303.07263\/} (2023).

\bibitem{winnie}
{\sc Jung, J., Tong, S., Hu, H., Lim, J., Jin, Y., and Kim, T.}
\newblock Winnie: Fuzzing windows applications with harness synthesis and fast cloning.
\newblock In {\em Proceedings of the 2021 Network and Distributed System Security Symposium (NDSS 2021)\/} (2021).

\bibitem{kang2021active}
{\sc Kang, H.~J., and Lo, D.}
\newblock Active learning of discriminative subgraph patterns for api misuse detection.
\newblock {\em IEEE Transactions on Software Engineering 48}, 8 (2021), 2761--2783.

\bibitem{Exception}
{\sc Kechagia, M., Devroey, X., Panichella, A., Gousios, G., and van Deursen, A.}
\newblock Effective and efficient api misuse detection via exception propagation and search-based testing.
\newblock In {\em Proceedings of the 28th ACM SIGSOFT International Symposium on Software Testing and Analysis\/} (2019), p.~192–203.

\bibitem{zero_cot}
{\sc Kojima, T., Gu, S.~S., Reid, M., Matsuo, Y., and Iwasawa, Y.}
\newblock Large language models are zero-shot reasoners.
\newblock In {\em Advances in Neural Information Processing Systems 35 (NIPS 2022)\/} (2022).

\bibitem{CodaMosa}
{\sc Lemieux, C., Inala, J.~P., Lahiri, S.~K., and Sen, S.}
\newblock Codamosa: Escaping coverage plateaus in test generation with pre-trained large language models.
\newblock In {\em 2023 IEEE/ACM 45th International Conference on Software Engineering (ICSE)\/} (2023), pp.~919--931.

\bibitem{lewis2020retrieval}
{\sc Lewis, P., Perez, E., Piktus, A., Petroni, F., Karpukhin, V., Goyal, N., K{\"u}ttler, H., Lewis, M., Yih, W.-t., Rockt{\"a}schel, T., et~al.}
\newblock Retrieval-augmented generation for knowledge-intensive nlp tasks.
\newblock In {\em Advances in Neural Information Processing Systems\/} (2020), vol.~33, pp.~9459--9474.

\bibitem{evalplus}
{\sc Liu, J., Xia, C.~S., Wang, Y., and Zhang, L.}
\newblock Is your code generated by chatgpt really correct? rigorous evaluation of large language models for code generation.
\newblock {\em arXiv preprint arXiv:2305.01210\/} (2023).

\bibitem{avatar}
{\sc Liu, K., Koyuncu, A., Kim, D., and Bissyandè, T.~F.}
\newblock Avatar: Fixing semantic bugs with fix patterns of static analysis violations.
\newblock In {\em 2019 IEEE 26th International Conference on Software Analysis, Evolution and Reengineering (SANER)\/} (2019), pp.~1--12.

\bibitem{afgen}
{\sc Liu, Y., Wang, Y., Bao, T., Jia, X., Zhang, Z., and Su, P.}
\newblock Afgen: Whole-function fuzzing for applications and libraries.
\newblock In {\em 2024 IEEE Symposium on Security and Privacy (SP)\/} (2024), pp.~11--11.

\bibitem{fdp}
{\sc LLVM}.
\newblock Fuzzeddataprovider.
\newblock \url{https://github.com/llvm/llvm-project/blob/main/compiler-rt/include/fuzzer/FuzzedDataProvider.h}.

\bibitem{libfuzzer}
{\sc LLVM}.
\newblock libfuzzer – a library for coverage-guided fuzz testing.
\newblock \url{https://llvm.org/docs/LibFuzzer.html}.

\bibitem{ubsan}
{\sc LLVM}.
\newblock Undefinedbehaviorsanitizer.
\newblock \url{https://clang.llvm.org/docs/UndefinedBehaviorSanitizer.html}.

\bibitem{lv2020rtfm}
{\sc Lv, T., Li, R., Yang, Y., Chen, K., Liao, X., Wang, X., Hu, P., and Xing, L.}
\newblock Rtfm! automatic assumption discovery and verification derivation from library document for api misuse detection.
\newblock In {\em Proceedings of the 2020 ACM SIGSAC conference on computer and communications security\/} (2020), pp.~1837--1852.

\bibitem{Goshawk}
{\sc Lyu, Y., Fang, Y., Zhang, Y., Sun, Q., Ma, S., Bertino, E., Lu, K., and Li, J.}
\newblock Goshawk: Hunting memory corruptions via structure-aware and object-centric memory operation synopsis.
\newblock In {\em 2022 2022 IEEE Symposium on Security and Privacy (SP) (SP)\/} (Los Alamitos, CA, USA, may 2022), IEEE Computer Society, pp.~1566--1566.

\bibitem{SparrowHawk}
{\sc Lyu, Y., Gao, W., Ma, S., Sun, Q., and Li, J.}
\newblock Sparrowhawk: Memory safety flaw detection via data-driven source code annotation.
\newblock In {\em Information Security and Cryptology: 17th International Conference, Inscrypt 2021, Virtual Event, August 12–14, 2021, Revised Selected Papers\/} (Berlin, Heidelberg, 2021), Springer-Verlag, p.~129–148.

\bibitem{nielebock2022automated}
{\sc Nielebock, S., Blockhaus, P., Kr{\"u}ger, J., and Ortmeier, F.}
\newblock Automated change rule inference for distance-based api misuse detection.
\newblock {\em arXiv preprint arXiv:2207.06665\/} (2022).

\bibitem{correction}
{\sc Nielebock, S., Heum\"{u}ller, R., Kr\"{u}ger, J., and Ortmeier, F.}
\newblock Cooperative api misuse detection using correction rules.
\newblock In {\em Proceedings of the ACM/IEEE 42nd International Conference on Software Engineering: New Ideas and Emerging Results\/} (2020), p.~73–76.

\bibitem{openai_chat}
{\sc OPENAI}.
\newblock How to use chat-based language models.
\newblock \url{https://platform.openai.com/docs/guides/chat/introduction}.

\bibitem{GPT3.5}
{\sc OPENAI}.
\newblock Introducing chatgpt.
\newblock \url{https://openai.com/blog/chatgpt}.

\bibitem{gpt4}
{\sc OpenAI}.
\newblock Gpt-4 technical report.
\newblock {\em arXiv preprint arXiv:2303.08774\/} (2023).

\bibitem{instruct_gpt}
{\sc Ouyang, L., Wu, J., Jiang, X., Almeida, D., Wainwright, C., Mishkin, P., Zhang, C., Agarwal, S., et~al.}
\newblock Training language models to follow instructions with human feedback.
\newblock In {\em Advances in Neural Information Processing Systems\/} (2022), pp.~27730--27744.

\bibitem{asleep}
{\sc Pearce, H., Ahmad, B., Tan, B., Dolan-Gavitt, B., and Karri, R.}
\newblock Asleep at the keyboard? assessing the security of github copilot’s code contributions.
\newblock In {\em 2022 IEEE Symposium on Security and Privacy (SP)\/} (2022), IEEE, pp.~754--768.

\bibitem{peng2023check}
{\sc Peng, B., Galley, M., He, P., Cheng, H., Xie, Y., Hu, Y., Huang, Q., Liden, L., Yu, Z., Chen, W., and Gao, J.}
\newblock Check your facts and try again: Improving large language models with external knowledge and automated feedback.
\newblock {\em arXiv preprint arXiv:2302.12813\/} (2023).

\bibitem{gpt2}
{\sc Radford, A., Wu, J., Child, R., Luan, D., Amodei, D., Sutskever, I., et~al.}
\newblock Language models are unsupervised multitask learners.
\newblock {\em OpenAI blog 1}, 8 (2019), 9.

\bibitem{lost_at_c}
{\sc Sandoval, G., Pearce, H., Nys, T., Karri, R., Garg, S., and Dolan-Gavitt, B.}
\newblock Lost at c: A user study on the security implications of large language model code assistants.
\newblock In {\em 32th USENIX Security Symposium (USENIX Security 23)\/} (2023).

\bibitem{oss_fuzz}
{\sc Serebryany, K.}
\newblock {OSS-Fuzz}-google's continuous fuzzing service for open source software.
\newblock In {\em Proceedings of the 26th USENIX Conference on Security Symposium (technical sessions)\/} (2017), USENIX Association.

\bibitem{asan}
{\sc Serebryany, K., Bruening, D., Potapenko, A., and Vyukov, D.}
\newblock Addresssanitizer: A fast address sanity checker.
\newblock In {\em Proceedings of the 2012 USENIX Conference on Annual Technical Conference\/} (2012), USENIX ATC'12, USENIX Association, p.~28.

\bibitem{ernie}
{\sc Sun, Y., Wang, S., Feng, S., et~al.}
\newblock Ernie 3.0: Large-scale knowledge enhanced pre-training for language understanding and generation.
\newblock {\em arXiv preprint arXiv:2107.02137\/} (2021).

\bibitem{clang_ast}
{\sc Tolnay, D.}
\newblock Clang ast deserializer in rust.
\newblock \url{https://github.com/dtolnay/clang-ast}.

\bibitem{llama}
{\sc Touvron, H., Martin, L., Stone, K., et~al.}
\newblock Llama 2: Open foundation and fine-tuned chat models.
\newblock {\em arXiv preprint arXiv:2307.09288\/} (2023).

\bibitem{carpetfuzz}
{\sc Wang, D., Li, Y., Zhang, Z., and Chen, K.}
\newblock Carpetfuzz: Automatic program option constraint extraction from documentation for fuzzing.
\newblock In {\em Proceedings of the 32nd USENIX Conference on Security Symposium\/} (Anaheim, CA, USA, 2023), USENIX Association.

\bibitem{superion}
{\sc Wang, J., Chen, B., Wei, L., and Liu, Y.}
\newblock Superion: Grammar-aware greybox fuzzing.
\newblock In {\em 2019 IEEE/ACM 41st International Conference on Software Engineering (ICSE)\/} (2019).

\bibitem{mutapi}
{\sc Wen, M., Liu, Y., Wu, R., Xie, X., Cheung, S.-C., and Su, Z.}
\newblock Exposing library api misuses via mutation analysis.
\newblock In {\em 2019 IEEE/ACM 41st International Conference on Software Engineering (ICSE)\/} (2019), pp.~866--877.

\bibitem{xia2023universal}
{\sc Xia, C.~S., Paltenghi, M., Tian, J.~L., Pradel, M., and Zhang, L.}
\newblock Universal fuzzing via large language models.
\newblock {\em arXiv preprint arXiv:2308.04748\/} (2023).

\bibitem{xia2023automated}
{\sc Xia, C.~S., Wei, Y., and Zhang, L.}
\newblock Automated program repair in the era of large pre-trained language models.
\newblock In {\em Proceedings of the 45th International Conference on Software Engineering (ICSE 2023). Association for Computing Machinery\/} (2023).

\bibitem{xia2023conversational}
{\sc Xia, C.~S., and Zhang, L.}
\newblock Conversational automated program repair.
\newblock {\em arXiv preprint arXiv:2301.13246\/} (2023).

\bibitem{yang2022api}
{\sc Yang, J., Ren, J., and Wu, W.}
\newblock Api misuse detection method based on transformer.
\newblock In {\em 2022 IEEE 22nd International Conference on Software Quality, Reliability and Security (QRS)\/} (2022), IEEE, pp.~958--969.

\bibitem{promptfuzz}
{\sc Yunlong~Lyu, Yuxuan~Xie, P.~C., and Chen, H.}
\newblock Promptfuzz.
\newblock \url{https://github.com/PromptFuzz/PromptFuzz}.

\bibitem{afl}
{\sc Zalewski, M.}
\newblock American fuzzy lop.
\newblock \url{http://lcamtuf.coredump.cx/afl/}.

\bibitem{zeng2021mining}
{\sc Zeng, H., Chen, J., Shen, B., and Zhong, H.}
\newblock Mining api constraints from library and client to detect api misuses.
\newblock In {\em 2021 28th Asia-Pacific Software Engineering Conference (APSEC)\/} (2021), IEEE, pp.~161--170.

\bibitem{zhang2023understanding}
{\sc Zhang, C., Bai, M., Zheng, Y., Li, Y., Xie, X., Li, Y., Ma, W., Sun, L., and Liu, Y.}
\newblock Understanding large language model based fuzz driver generation.
\newblock {\em arXiv preprint arXiv:2307.12469\/} (2023).

\bibitem{apicraft}
{\sc Zhang, C., Lin, X., Li, Y., Xue, Y., Xie, J., Chen, H., Ying, X., Wang, J., and Liu, Y.}
\newblock {APICraft}: Fuzz driver generation for closed-source {SDK} libraries.
\newblock In {\em 30th USENIX Security Symposium (USENIX Security 21)\/} (2021), pp.~2811--2828.

\bibitem{intelligen}
{\sc Zhang, M., Liu, J., Ma, F., Zhang, H., and Jiang, Y.}
\newblock Intelligen: Automatic driver synthesis for fuzz testing.
\newblock In {\em 2021 IEEE/ACM 43rd International Conference on Software Engineering: Software Engineering in Practice (ICSE-SEIP)\/} (2021), IEEE, pp.~318--327.

\bibitem{Zhao:2023}
{\sc Zhao, J., Rong, Y., Guo, Y., He, Y., and Chen, H.}
\newblock Understanding programs by exploiting (fuzzing) test cases.
\newblock In {\em Findings of the Association for Computational Linguistics (ACL)\/} (Toronto, Canada, 2023).

\end{thebibliography}
\appendix

\section{Example of converted fuzz driver.}
\autoref{fig: example_convert} shows a case of constrained argument conversion described in \autoref{sec:conversion}.

\begin{figure}[!htb]
\begin{lstlisting}[language=C]
#include "FuzzedDataProvider.h"
#include <cJSON.h>

extern "C" int LLVMFuzzerTestOneInput_14(const uint8_t* f_data, size_t f_size) { 
//fuzzer vars shim {
    FuzzedDataProvider fdp(f_data, f_size);
    FDPConsumeRawBytesWithNullTerm(const uint8_t *, data, size, fdp)
    FDPConsumeRandomLengthString(char, fuzz_str_1, fuzz_str_sz_1, fdp);
    FDPConsumeFloatingPoint(double, fuzz_var_2, fdp)
    FDPConsumeFloatingArray(float, fuzz_array_3, fuzz_array_size_3, fdp);
//fuzzer shim end}

    // Parse the JSON data
    cJSON *root = cJSON_ParseWithLength(data, size);
    
    // Add a number value to the root object
(*@\colorbox{lightgray}{\parbox{0.98\linewidth}{- cJSON\_AddNumberToObject(root, "pi", 3.14);}}@*)
(*@\colorbox{Melon}{\parbox{0.98\linewidth}{+ cJSON\_AddNumberToObject(root, fuzz\_str\_1, fuzz\_var\_2);}}@*)

    // Create a float array and add it to the root object
(*@\colorbox{lightgray}{\parbox{0.98\linewidth}{- cJSON *array = cJSON\_CreateFloatArray("{1.23f, 4.56f, 7.89f}", 3);}}@*)
(*@\colorbox{Melon}{\parbox{0.98\linewidth}{+cJSON *array = cJSON\_CreateFloatArray(fuzz\_array\_3, fuzz\_array\_size\_3);}}@*)
    ...
    // Delete the cJSON object
    cJSON_Delete(root);
    return 0;
}
\end{lstlisting}
\caption{An example of constrained argument conversion.}
\label{fig: example_convert}
\Description[conversion example]{An example of constrained argument conversion.}
\end{figure}

\section{Examples of Library Specification.}

\begin{figure}[!htb]
\begin{lstlisting}[language=C]
// Generate the fuzz driver with the beginning code:
extern "C" int LLVMFuzzerTestOneInput(const uint8_t* data, size_t size) {
    magic_t magic = magic_open(MAGIC_NONE);
    if (magic == NULL) {
        return -1;
    }
    // The magic file name is "magic"
    if (magic_load(magic, "magic") == -1) {
        magic_close(magic);
        return -1;
    }
  // complete here
\end{lstlisting}
\caption{The library specification used for libmagic.}
\label{fig: spec_libmagic}
\Description[libmagic specification]{The library specification used for libmagic.}
\end{figure}

\begin{figure}[!htb]
\begin{lstlisting}[language=C]
// Generate the fuzz driver with the beginning code:
extern "C" int LLVMFuzzerTestOneInput(const uint8_t* data, size_t size) {
    sqlite3* db;
    int rc = sqlite3_open(":memory:", &db);
    if (rc != SQLITE_OK) {
      sqlite3_close(db); 
      return 0;
    }
    // complete here
\end{lstlisting}
\caption{The library specification used for sqlite3.}
\label{fig: spec_sqlite3}
\Description[sqlite3 specification]{The library specification used for sqlite3.}
\end{figure}

\begin{figure}[!htb]
\begin{lstlisting}[language=C]
// Generate the fuzz driver with the beginning code:
extern "C" int LLVMFuzzerTestOneInput(const uint8_t* data, size_t size) {
    // write data into input_file.
    FILE *in_file = fopen("input_file", "wb");
    if (in_file == NULL) {return 0;}
    fwrite(data, sizeof(uint8_t), size, in_file);
    fclose(in_file);
    // open input tiff in memory
    std::istringstream s(std::string(data, data + size));
    TIFF *in_tif = TIFFStreamOpen("MemTIFF", &s);
    if (!in_tif)
    {
      return 0;
    }
    // complete here
\end{lstlisting}
\caption{The library specification used for libTIFF.}
\label{fig: spec_libtiff}
\Description[libtiff specification]{The library specification used for libTIFF.}
\end{figure}

As stated in \autoref{sec: generation}, we can provide the library specification to LLMs if the library requires specialized guidelines.
The provided library specifications are filled in our prompt template (\autoref{fig: prompt}) and used to facilitate the specific code pattern generation.
Among the 14 tested libraries in our experiments,
    we manually crafted the library specifications for 3 libraries: \textsf{libmagic}, \textsf{sqlite3}, and \textsf{libTIFF}.
The remaining 11 libraries were not crafted with additional library specifications.

\autoref{fig: spec_libmagic} shows the library specification for \textsf{libmagic}. 
The library \textsf{libmagic} requires a pre-loaded magic database as an argument to calling the \textsf{libmagic} API functions. To initialize this, the file location of the magic database file needs to be passed to the function \func{magic\_load()}. To ensure that LLMs can successfully load the magic database, we explicitly specified the file name of the magic database and prepared it in the correct location in advance.

The library \textsf{sqlite3} requires actively calling \func{sqlite3\_close()} to close the database opened by \func{sqlite3\_open()}, regardless of whether the database is opened successfully. This code pattern against the common pattern of \func{*\_open()} and \func{*\_close()} pairs. Memory leaks will occur if the API \func{sqlite3\_close()} has not been called to close the opened database. We crafted the library specification in \autoref{fig: spec_sqlite3} to mitigate the memory leak issues.

The library \textsf{libTIFF} provides multiple methods for opening TIFF (Tag Image File Format) files. For instance, \func{TIFFOpen()} opens a TIFF file using a file location, \func{TIFFFdOpen()} opens a TIFF file using a file descriptor, and \func{TIFFStreamOpen()} opens a TIFF file using a byte stream. In our experiments, LLMs tend to utilize \func{fmemopen()} to obtain a file descriptor and then use \func{TIFFFdOpen()} to open the TIFF file. However, if the file descriptor is obtained through \func{fmemopen()}, \func{TIFFFdOpen()} will consistently fail. We designed the library specification presented in \autoref{fig: spec_libtiff} to ensure the correct code pattern is generated.

\end{document}